\documentclass[fleqn,usenatbib]{mnras}
\usepackage[T1]{fontenc}
\DeclareRobustCommand{\VAN}[3]{#2}
\let\VANthebibliography\thebibliography
\def\thebibliography{\DeclareRobustCommand{\VAN}[3]{##3}\VANthebibliography}

\usepackage{graphicx}	
\usepackage{amsmath}
\usepackage{amssymb}
\usepackage{color}
\usepackage{adjustbox}

\usepackage{threeparttable,longtable}
\usepackage{tikz,hyperref}
\usepackage{academicons}
\definecolor{lime}{HTML}{A6CE39}
\DeclareRobustCommand{\orcidicon}{
	\begin{tikzpicture}
	\draw[lime, fill=lime] (0,0) 
	circle [radius=0.16] 
	node[white] {{\fontfamily{qag}\selectfont \tiny ID}};
	\draw[white, fill=white] (-0.0625,0.095) 
	circle [radius=0.007];
	\end{tikzpicture}
	\hspace{-2mm}
}

\foreach \x in {A, ..., Z}{
	\expandafter\xdef\csname orcid\x\endcsname{\noexpand\href{https://orcid.org/\csname orcidauthor\x\endcsname}{\noexpand\orcidicon}}
}

\newcommand{\orcid}[1]{\href{https://orcid.org/#1}{\textcolor[HTML]{A6CE39}{\orcidicon}}}

\definecolor{frenchblue}{rgb}{0.0, 0.45, 0.73}
\definecolor{burgundy}{rgb}{0.5, 0.0, 0.13}
\definecolor{royalblue}{RGB}{65,105,225}
\definecolor{darkspringgreen}{rgb}{0.09, 0.45, 0.27}
\definecolor{pgcol}{rgb}{0.67, 0.16, 0.16}
\definecolor{mypink3}{cmyk}{0, 0.7808, 0.4429, 0.1412}

\graphicspath{{./}{plots/}{figures/}}
\newcommand{\swift}{\emph{Swift}}
\newcommand{\icnu}{IceCube-170922A}
\newcommand{\txs}{TXS~0506+056}
\title[Neutrinos from X-ray flares]{High-energy neutrinos from X-rays flares of blazars frequently observed by the \emph{Swift} X-Ray Telescope
}
\author[Stamatios~I. Stathopoulos et al.]{S.~I. Stathopoulos$^{1}$\thanks{E-mail: stamstath@yahoo.gr},
M.~Petropoulou\orcid{0000-0001-6640-0179}$^{1}$\thanks{Mercator Fellow},
P.~Giommi\orcid{0000-0002-2265-5003}$^{2,3,4}$,
G.~Vasilopoulos\orcid{0000-0003-3902-3915}$^{5,6}$,
\newauthor 
P.~Padovani\orcid{0000-0002-4707-6841}$^{7}$,
A.~Mastichiadis\orcid{0000-0001-5217-4801}$^1$
\\
$^{1}$Department of Physics, National and Kapodistrian University of Athens, University Campus Zografos, GR 15783, Greece\\
$^{2}$ Institute for Advanced Study, Technische Universität München, Lichtenbergstrasse 2a, 85748, Garching bei München, Germany \\
$^3$Associated to the Italian Space Agency, ASI, Via del Politecnico snc, 00133, Roma, Italy \\
$^{4}$Center for Astro, Particle and Planetary Physics, New York University Abu Dhabi\\
$^5$Department of Astronomy, Yale University, PO Box 208101, New Haven, CT 06520-8101, USA  \\
$^6$Université de Strasbourg, CNRS, Observatoire astronomique de Strasbourg, UMR 7550, 67000, Strasbourg, France \\
$^{7}$European Southern Observatory, Karl-Schwarzschild-Str. 2, Garching bei M{\"u}nchen D-85748, Germany \\
}

\begin{document}
\label{firstpage}
\pagerange{\pageref{firstpage}--\pageref{lastpage}}
\maketitle
% Abstract of the paper

\begin{abstract}
Blazar flares have been suggested as
ideal candidates for enhanced neutrino production. While the neutrino signal of $\gamma$-ray flares has been widely discussed, the neutrino yield of X-ray flares has received less attention. Here, we
compute the predicted neutrino signal from X-ray flares detected in 66 blazars observed  more than 50 times with the X-ray Telescope (XRT) on board the  Neil Gehrels \swift \, Observatory. We consider a scenario where X-ray flares are powered by synchrotron radiation of relativistic protons, and neutrinos are produced through photomeson interactions between protons with their own synchrotron X-ray photons. Using the 1~keV X-ray light curves for flare identification, the 0.5-10~keV fluence of each flare as a proxy for the all-flavour neutrino fluence, and the  IceCube point-source effective area for different detector configurations, we calculate the number of muon and antimuon neutrinos above 100~TeV expected for IceCube from each flaring source. The bulk of the neutrino events from the sample originates from flares with durations $\sim 1-10$~d. Accounting for the X-ray flare duty cycle of the sources in the sample,  which ranges between $\sim2$ and 24 per cent, we compute an average yearly neutrino rate for each source. The median of the distribution (in logarithm) is $\sim0.03$~yr$^{-1}$, with Mkn~421 having the highest predicted rate $1.2\pm 0.3$~yr$^{-1}$, followed by 3C~273 ($0.33\pm0.03$~yr$^{-1}$) and PG~1553+113 ($0.25\pm0.02$~yr$^{-1}$). Next-generation neutrino detectors together with regular X-ray monitoring of blazars could constrain the duty cycle of hadronic X-ray flares.
\end{abstract}

\begin{keywords}
 galaxies: active -- neutrinos -- radiation mechanisms: non-thermal  --  X-rays: galaxies 
\end{keywords}

\section{Introduction}
Blazars are a subclass of active galactic nuclei (AGN) with relativistic jets closely aligned to our line of sight \citep[e.g.][]{Urry1995} which are powered by accretion onto a central supermassive black hole \citep[e.g.][]{Begelman1984}. They are the most powerful persistent astrophysical sources of non-thermal electromagnetic radiation in the Universe, with spectral energy distributions (SEDs) spanning $\sim15$ decades in energy, from radio frequencies up to high-energy $\gamma$-rays \citep[for a recent review on AGN, see][]{Padovani_2017}. 

The blazar SED has a characteristic double-humped appearance (in a $\varepsilon F({\varepsilon})$ space) with the low-energy component peaking between infrared and X-ray energies and the high-energy component peaking in $\gamma$-rays \citep[e.g.][]{Ulrich_1997, Fossati_1998}. The low-energy hump is generally attributed to synchrotron radiation produced in a localized region of the jet (aka a blob) by a population of relativistic electrons. The origin of the high-energy component, however, is less clear, with two alternative scenarios put forward to explain it. In leptonic scenarios, high-energy photons are produced via inverse Compton scattering between relativistic electrons in the jet and their own synchrotron photons \citep[synchrotron self-Compton, see e.g. ][]{1992ApJ...397L...5M, 1996ApJ...461..657B,  1997A&A...320...19M} or low-energy external radiation fields \citep[external Compton, see e.g.][]{1992A&A...256L..27D,  1994ApJ...421..153S, 1996MNRAS.280...67G}. In hadronic scenarios, high-energy emission is either explained by synchrotron radiation of relativistic protons \citep{2000NewA....5..377A, 2001APh....15..121M} or by synchrotron (or inverse Compton) processes of secondary electrons and positrons produced from proton-photon interactions and photon-photon pair production in the jet \citep[e.g.][]{1991A&A...251..723M,1991PhRvL..66.2697S,  1993A&A...269...67M, 2003APh....18..593M}. The latter class of models also predicts high-energy  muon and electron neutrinos from the decay of charged pions produced in photomeson interactions. Hence, the detection of high-energy neutrinos from individual blazars would be the smoking gun of baryon-loaded jets acting as cosmic ray accelerators.

In 2013 the IceCube neutrino telescope discovered a diffuse flux of astrophysical neutrinos at energies exceeding a few tens of TeV \citep{Aartsen:2013a, Aartsen:2013b}. The absence of a significant anisotropy is consistent with the majority ($\gtrsim 85$ per cent) of the neutrino signal coming from extragalactic sources \citep[e.g.][]{Aartsen_2017,Aartsen_2019}. Various astrophysical populations have been suggested to explain the diffuse flux observed by IceCube~\citep[for a recent review, see][]{2017ARNPS..67...45M}. Even though the identity of the sources producing the diffuse flux remains largely unknown, strong constraints have already been placed on specific classes
by the lack of correlations between high-energy neutrinos and known sources or the lack of significant clustering in high-energy neutrino events \citep[e.g.][]{Murase_2016, Aartsen_2017_grb,2020ApJ...890...25Y}.   

Searches for transient electromagnetic phenomena, such as blazar flares, could improve the association of neutrinos with astrophysical point sources, since both the arrival time and direction of the detected events could be utilized, while the contribution from the atmospheric background could be much smaller than the signal.\footnote{Bright flaring sources are detectable in neutrinos regardless of the contribution of the blazar population to   the extragalactic neutrino sky \citep[e.g.][]{Murase_2016, Guepin_2017,Murase_2018}.}. Such a multi-messenger approach led to the first association in time and space  of a high-energy neutrino event, \icnu, with a $\gamma$-ray flaring blazar  \txs~(at the $\sim3\sigma$ level) \citep{IceCube:2018dnn}. 
A follow-up archival search of more than 9 years of IceCube data revealed an excess of high-energy neutrinos with respect to the atmospheric background over a period of $\sim 6$ months in 2014-2015. This finding provided a $\sim 3.5\sigma$ evidence for neutrino emission from the direction of \txs~\citep{IceCube:2018cha}. Notably, during that time  the  source  was not flaring at any wavelength (from radio up to GeV $\gamma$-rays)~\citep{IceCube:2018cha, 2019ApJ...880..103G}.
 
From a theoretical perspective, assuming a hadronic scenario, periods of flaring activity are considered to be ideal for enhancing the predicted neutrino signal, as long as both messengers (photons and neutrinos) are produced at the same site. The increased electromagnetic flux during flares usually implies that the density of photons used as targets for photomeson interactions with relativistic protons in the blazar jet is higher and/or the injection rate of accelerated protons is enhanced. As a result, many models predict that the all-flavour neutrino luminosity, $L_{\nu}$, is strongly enhanced during flares, with $L_{\nu} \propto L_{\rm ph}^{\gamma}$, where $L_{\rm ph}$ is the photon luminosity in some energy band and $\gamma \sim 1.5 - 2$~\citep[e.g.][]{Murase_2014,Tavecchio_2014, Petropoulou_2016, Murase_2018}.

The Large Area Telescope (LAT) on board the \textit{Fermi} Gamma-Ray Space Telescope~\citep[][]{LAT} has been instrumental in searches of $\gamma$-ray electromagnetic counterparts to IceCube high-energy neutrinos~\citep[e.g.][]{Brown_2015, Padovani_2016, Palladino_2017, Murase_2018, 2020MNRAS.497..865G, 2021JCAP...03..031S}. With an $\sim13$ year-long operation period, \textit{Fermi}-LAT produced a large sample of long-term blazar $\gamma$-ray light curves with regular sampling that also enables correlation studies of $\gamma$-ray flares and high-energy neutrinos \citep{Oikonomou_2019, 2019ICRC...36.1038Y, 2020ApJ...893..162F}. The discovery potential of these searches, however, depends strongly on the intrinsic opacity of the source in $\gamma \gamma$ pair production at GeV energies. GeV $\gamma$-ray dark sources could still be bright neutrino emitters \citep[e.g.][]{2016PhRvL.116g1101M}, but would be missed by \textit{Fermi}-LAT searches. For instance, the lack of flaring activity in GeV $\gamma$-rays during the period of the neutrino excess in \txs~\citep{IceCube:2018cha} -- assuming that the detected neutrinos are truly of astrophysical origin -- suggests attenuation of multi-GeV photons on low-energy photons~\citep[e.g.][]{Reimer_2019, Rodrigues_2019, 2020ApJ...891..115P} or decoupled regions for GeV photon and PeV neutrino production \citep[e.g.][]{2019ApJ...886...23X,2020ApJ...889..118Z}. 

Motivated by the possibility that high-energy neutrinos are not always correlated with $\gamma$-ray flares, \cite{2021ApJ...906..131M} presented an alternative scenario that relates X-ray flares with TeV-PeV neutrinos. In their model, X-ray flares occur whenever protons are accelerated intermittently to high enough energies in the blazar jet, and produce pions interacting mainly with proton-synchrotron radiation. The reason for focusing on X-ray flares is twofold: X-ray photons are energetic targets for photomeson interactions, thus reducing the required proton energy for pion production and, at the same time, can be plentiful providing substantial optical thickness for the interactions. Notably, the X-ray flux of the proposed hadronic flares is a good proxy for the all-flavour neutrino flux, while certain neutrino-rich X-ray flares may be dark in GeV-TeV $\gamma$-rays.

In this work, we present quantitative neutrino predictions of the hadronic X-ray flaring scenario of blazars. We compute the number of muon and antimuon neutrinos above 100~TeV expected for IceCube from X-ray flares of blazars that were observed  more than 50 times with the X-ray Telescope  (XRT, \citealt{xrt2005}) on board the Neil Gehrels \emph{Swift} Observatory between November 2004 and November 2020 \citep{2021arXiv210807255G}. To this end, we apply the Bayesian block algorithm to the 1~keV XRT light curves of these frequently observed blazars to characterize statistically significant variations and identify flares. Using X-ray spectral
information in the 0.5-10 keV energy range, and the duration of each flaring block as a proxy of the flare duration, we compute the all-flavour neutrino fluence of each flare. Adopting the point-source effective area of IceCube, we compute the predicted number of muon and antimuon neutrinos per flare from each source. To the best of our knowledge, this is the first time that \swift/XRT data are utilized for this purpose. 

This paper is structured as follows. In Section~\ref{sec:model} we summarize the theoretical model, highlighting the ingredients needed for the estimation of neutrino events from X-ray flares. In Section~\ref{sec:data} we present the dataset and methods used for the search of X-ray flares. In Section~\ref{sec:neutrino-counts} we present our method for computing the expected number of muon and antimuon neutrino events from X-ray flares. We continue with a presentation of our results in Section~\ref{sec:results}. 
We conclude with a summary and discussion of our findings in Section~\ref{sec:discussion}.

\section{Theoretical Model}\label{sec:model}
The basic assumption of our model 
is that every X-ray blazar flare 
is produced by the synchrotron radiation of a hadronic (proton) population that is intermittently accelerated in the blazar jet. The non-flaring X-ray emission of the source does not necessarily originate from the same region as the flares. In our scenario we assume that it is attributed to radiative processes of electrons accelerated in the blazar jet and will not be discussed further. 

Upon acceleration to relativistic energies, protons are assumed to be injected into a region where they lose energy via radiative processes, including synchrotron radiation and photomeson interactions. In particular, photomeson interactions of protons with their own synchrotron photons lead to the production of a simultaneous high-energy neutrino flare. 
From the peak frequency of the X-ray flare spectrum we can infer the minimum proton energy needed to produce neutrinos through photomeson interactions with proton synchrotron photons. In addition, we relate the flux of the X-ray flare to the all-flavour neutrino and anti-neutrino flux through a theoretically motivated scaling parameter $\xi_{\rm X}$. Detailed numerical calculations of the broadband photon spectra and the associated neutrino emission in the proposed scenario for X-ray blazar flares have been presented in \cite{2021ApJ...906..131M}. For completeness, we briefly discuss the model ingredients that are also necessary for this work.

Relativistic protons with Lorentz factor $\gamma'_{\rm p}$ in the presence of a magnetic field with strength $B'$ radiate synchrotron photons of characteristic observed energy $\varepsilon$. Henceforth, primed quantities are measured in the rest frame of the emission region, while unprimed quantities correspond to the measurements in the observer's frame. The proton Lorentz factor can be written as
\begin{equation}
\gamma'_{\rm p}\simeq 1.4\times10^6\sqrt{\mathcal{D}_1^{-1}B_1^{'-1} \varepsilon_{\rm keV}(1+z)}
\label{eq:gp}
\end{equation}
where $\mathcal{D}$ is the Doppler factor that corresponds to the relativistic motion of the emission region, $z$ the redshift of the source, $\varepsilon_{\rm keV}=\varepsilon/1~{\rm keV}$ and the notation $q_x = q/10^x$ in cgs units was introduced, unless stated otherwise.

Protons with Lorentz factors given by equation~(\ref{eq:gp}) would produce neutrinos if they exceed the energy threshold for photomeson interactions with synchrotron photons of energy $\varepsilon$. This translates to a minimum proton Lorentz factor given by
 \begin{equation}
    \gamma'_{\rm p,th}=1.4\times 10^6(1 + z)^{-1}\mathcal{D}_1\varepsilon_{\rm keV}^{-1}
	\label{eq:gpth}
\end{equation}

Provided that $\gamma_{\rm p}^\prime \gtrsim  \gamma'_{\rm p,th}$, the energy of neutrinos produced by protons radiating synchrotron photons of energy $\varepsilon$ is 
\begin{eqnarray}
\label{eq:Enu}
\varepsilon_{\nu} &\simeq& 0.05 \,  \mathcal{D}(1+z)^{-1}\gamma'_{\rm p}m_{\rm p} c^2 \\ \nonumber 
& \simeq & 0.6 \, \sqrt{\mathcal{D}_1 B_1^{'-1}\varepsilon_{\rm keV}(1+z)^{-1}}~{\rm PeV}
\end{eqnarray}
where we used equation~(\ref{eq:gp}).

Ignoring the Bethe-Heitler pair production as a cooling process for protons, the ratio of neutrino-to-photon luminosities can be written as 
 \begin{equation}
	\frac{L_{\nu+\bar{\nu}}}{L_{\rm ph}} \approx \frac{(1-\alpha)t'^{-1}_{\rm mes}}{t'^{-1}_{\rm syn}+\alpha t'^{-1}_{\rm mes}}
	\label{eq:ratio}
\end{equation}
where $\alpha \simeq 5/8$ and $L_{\rm \nu+\bar{\nu}}, L_{\rm ph}$ are the bolometric neutrino and photon luminosities. Here, we focus on the ``neutrino-rich'' scenario where the photomeson energy loss rate is comparable to the energy loss rate due to proton-synchrotron radiation $t'^{-1}_{\rm syn}\simeq t'^{-1}_{\rm mes}$ \citep[for details see][]{2021ApJ...906..131M}. In this case, equation~(\ref{eq:ratio}) yields
 \begin{equation}
	L_{\nu+\bar{\nu}} =  \frac{1-\alpha}{1+\alpha} L_{\rm ph}\simeq 0.23 L_{\rm ph}.
	\label{eq:ratio-1}
\end{equation}
Detailed numerical calculations of proton-synchrotron-powered X-ray flares have shown that equation~(\ref{eq:ratio-1}) is indeed a good proxy for the neutrino luminosity close to the peak time of the X-ray flare. If we replace the bolometric photon luminosity with the  $0.5-10$~keV X-ray luminosity of the flare, equation~(\ref{eq:ratio-1}) becomes $L_{\nu +\bar{\nu}} = \xi_{\rm X} L_{\rm X}$ with $\xi_{\rm X}\approx 1$. 

We model the differential neutrino plus anti-neutrino energy flux of all flavours as
\begin{equation}
F_{\nu+\bar{\nu}}(\varepsilon_{\nu},t) = F_0(t)\varepsilon_{\nu}^{-s(t)} e^{-\varepsilon_{\nu}/\varepsilon_{\nu, \rm c}(t)}
\label{eq:Fnu}
\end{equation}
where the parameters to be defined are $\varepsilon_{\nu, \rm c}$ (characteristic energy), $s$ (spectral slope), and $F_0$ (normalization). Numerical calculations presented in  \cite{2021ApJ...906..131M} show that the slope $s$ does not change much during the flare with an average value $\langle s\rangle \simeq-0.5$. This is the value we are going to adopt in our calculations. The characteristic neutrino energy is found from equation~(\ref{eq:Enu}) with  $\varepsilon$ corresponding to the peak energy of the flare spectrum ($\varepsilon_{\rm pk}$)\footnote{This should not be confused with the peak energy of the low-energy component of the SED $\varepsilon^{\rm S}_{\rm pk}$.}, which can in principle change during the flare. We can calculate the normalization factor $F_0$ using the following relation  
 \begin{equation}
	\int_{\varepsilon_{\nu,\min}}^{\varepsilon_{\nu,\max}} {\rm d}\varepsilon_{\nu}F_{\nu+\bar{\nu}}(\varepsilon_{\nu},t)=\xi_{\rm X} \int_{\varepsilon_{\min}}^{\varepsilon_{\max}} {\rm d}\varepsilon \, F_{\rm X}(\varepsilon,t)
	\label{eq:fluxes}
\end{equation}
where $\varepsilon_{\nu,\min}\approx 0$\footnote{In reality, this is set by the minimum energy of protons interacting with their own synchrotron photons. Because $\varepsilon_{\nu,\min} \ll \varepsilon_{\nu, \rm c}$ and $s<0$, we use $\varepsilon_{\nu,\min} \approx0$.}, $\varepsilon_{\nu,\max}=\infty$, $\varepsilon_{\min}=0.5$~keV, $\varepsilon_{\max}=10$~keV and $F_{\rm X}(\varepsilon, t)$ is the differential photon energy flux in X-rays. If the latter can be described by a power law with photon index $\Gamma$ between $\varepsilon_{\min}$ and $\varepsilon_{\max}$ at time $t$, namely
 \begin{equation}
F_{\rm X}(\varepsilon,t)=F_{\rm X,0}(t)\varepsilon^{-\Gamma+1}
	\label{eq:fX}
\end{equation}
the normalization factor of the neutrino flux, $F_0(t)$, can be expressed as 
 \begin{equation}
     F_0(t) =
    \begin{cases}
    \xi_{\rm X} \frac{F_{\rm X,0}(t)\left(\varepsilon_{\max}^{-\Gamma+2}-\varepsilon_{\min}^{-\Gamma+2}\right) \varepsilon_{\nu, \rm c}^{s-1}} {(-\Gamma+2)\int_0^{\infty}  {\rm d}x x^{-s}e^{-x}},
    & \text{if } \Gamma\neq 2,\\ \\
    \xi_{\rm X} \frac{F_{\rm X,0}(t)\ln\left(\varepsilon_{\max}/\varepsilon_{\min}\right) \varepsilon_{\nu, \rm c}^{s-1}} {(-\Gamma+2)\int_0^{\infty}  {\rm d}x x^{-s}e^{-x}},
    & \text{if } \Gamma=2.
    \end{cases}
	\label{eq:F0}
\end{equation} 

\begin{figure} 
\centering
\includegraphics[width=0.47\textwidth]{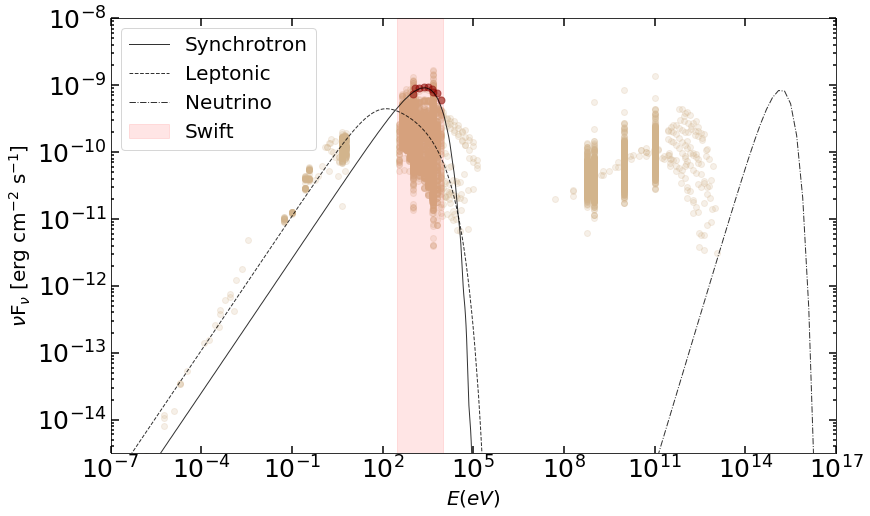}
\caption{Spectral energy distribution of Mkn~421 compiled using data from various instruments and epochs (adopted from the \href{http://openuniverse.asi.it/ouspectra/catalog_description.html}{Open Universe for Blazars}). The spectrum of an X-ray flare is highlighted with red symbols and the shaded region indicates the 0.5-10 keV energy range. Solid and dash-dotted lines present the proton synchrotron  spectrum and the accompanying all-flavour neutrino spectrum of the flare, respectively. A likely contribution to the non-flaring spectrum from an accelerated electron population is also displayed (dotted line).} 
\label{fig:sed}
\end{figure}

Our model is schematically shown in Fig.~\ref{fig:sed}. We present the proton synchrotron spectrum and the all-flavour neutrino spectrum (see equation~\ref{eq:Fnu}) from an X-ray flare of Mkn~421. In our model, proton synchrotron radiation is assumed to power the flaring X-ray states of a source, while other processes, such as electron synchrotron or inverse Compton scattering of low-energy photons, are responsible for the non-flaring X-ray emission. This leptonic component is schematically shown with the dotted line. Additional emission components accompanying the X-ray flare (from photopair and photopion processes) are not shown here \citep[for a complete treatment of the multi-wavelength emission, see][]{2021ApJ...906..131M}. 

\section{X-ray light curves}\label{sec:data}
We use data from the X-ray Telescope (XRT, \citealt{xrt2005}) on board the Neil Gehrels \emph{Swift} Observatory obtained between November 2004 and November 2020. Our sample comprises of
all blazars that have been observed at least 50 times in this period with \swift \, (see Table~\ref{tab:sample}). This amounts to 66 blazars out of which 26 are high-synchrotron peaked (HSPs) objects, 15 are intermediate-synchrotron peaked (ISPs) sources, and 25 are low-synchrotron peaked (LSPs) objects\footnote{Blazars are divided in spectral classes depending on the peak energy of the their low-energy (synchrotron) hump ($\varepsilon_{\rm pk}^{\rm S}$) into LSPs with $\varepsilon_{\rm pk}^{\rm S}<0.41$~eV, ISPs with 
$0.41<\varepsilon_{\rm pk}^{\rm S}<4.1$~eV, and HSPs with $\varepsilon_{\rm pk}^{\rm S}>4.1$~eV \citep{1995ApJ...444..567P,2010ApJ...716...30A}.}. 
We exclude 3 sources from our analysis (i.e. 1RXS~J154439.4-112820, 3HSP~J022539.1-190035, and 2E~1823.3+5649). For instance, 1RXS~J154439.4-112820 was pointed 55 times by \swift, but only 45 observations could be used for X-ray analysis. Typical reasons for excluding an observation  are very short exposures ($<200$~s) or a low count rate (e.g. the window-time readout mode is used when the source count rate is lower than 0.5 c/s, which makes spectral analysis unreliable). The case for 3HSP~J022539.1-190035 is different; this source was not the target of \swift \,  observations, but lied in the
field of view of GRB~091127, which was observed many times over a short time interval. Some of these sources (TXS~0506+056, 1ES~0229+200 /3HSP~J023248.5+20171 and PKS~1502+106/5BZQ~J1504+1029) have also been identified as possible counterparts of IceCube high-energy tracks \citep[e.g.][]{2016NatPh..12..807K, IceCube:2018cha,2019ApJ...880..103G, 2020ApJ...893..162F, 2020MNRAS.497..865G}. 

To search for X-ray variability we use the 1~keV X-ray light curves as obtained by \cite{2021arXiv210807255G}.
The \swift/XRT data products are based on the pipeline, the procedure, and methodology developed for the Open Universe for Blazars project \citep{2018arXiv180508505G,2019A&A...631A.116G}. Here, we provide a brief outline of the analysis procedure, but for a comprehensive description  we refer the reader to \cite{2021arXiv210807255G}. X-ray source and background events were extracted from XRT data. For data sets with enough counts (i.e. $>20$) X-ray fitting was performed with {\tt xspec} \citep{1996ASPC..101...17A} assuming an absorbed power-law model. The goodness of the fit was estimated using Cash statistics \citep{1979ApJ...228..939C}. From the best-fit model, the 1 keV fluxes were computed from the power-law normalization. For sources with less than 20 counts available, spectral fitting was not performed. Instead, count rates were estimated in different energy bands using source detection via an X-ray image package {\tt ximage}\footnote{Software is part of {\tt HEASOFT/FTOOLS} \url{http://heasarc.gsfc.nasa.gov/ftools}}, and the 1 keV fluxes were estimated by scaling the count rates and adopting generic parameters for the spectral model. To estimate the neutrino fluence of each X-ray flare, we use the integrated flux in the 0.5-10~keV energy range. This is either computed from the best spectral fit or by scaling the broadband XRT count rate, as described above.

Fig.~\ref{fig:lc} shows indicative 1~keV X-ray light curves from our sample (about the 0.5-10~keV light curves, see Appendix~\ref{app:lc}). Each point in the light curve is derived from individual XRT snapshots with typical duration of $\sim1$~ks. Most sources have sparse coverage in X-rays, despite belonging to the sample of frequently observed blazars with XRT, with observations clustered around times of interest. An illustrative example is the 2017 multi-wavelength flare of TXS~0506+056 that has been associated with the high-energy neutrino IC~170922A \citep{icecube2018a}. The lack of X-ray observations prior to that epoch makes difficult a detailed study of the long-term behaviour of this source in neutrinos \citep{2020ApJ...891..115P}.  
4FGL~J1544.3-0649 is a unique blazar, as it transitioned from being an anonymous mid-intensity radio source, never detected at high energies, to  one of the brightest extreme blazars ($\varepsilon^{\rm S}_{\rm pk}\gtrsim~1$~keV) in the sky \citep{2021MNRAS.502..836S}. It is one of the sources that would go unnoticed if the $\gamma$-ray intensity of the flare remained below the sensitivity of \textit{Fermi}-LAT or if the $\gamma$-ray emission was intrinsically low during the X-ray flare, as predicted in the hadronic scenario under study for certain parameters \citep{2021ApJ...906..131M}. Only a few sources, like Mkn~421 and Mkn~501, have well sampled light curves, thus allowing a robust characterization of their long-term X-ray variability properties. Large amplitude variability (i.e. changes in flux by a factor $\sim2-3$) on different timescales is clearly present in all sources displayed in Fig.~\ref{fig:lc}. X-ray flares are ubiquitous in the blazars in our sample. In the next paragraph we describe how we define X-ray flares whose properties   (i.e. duration and flux) are presented in Section~\ref{sec:flares-results}.

\begin{figure*}
\includegraphics[width=0.47\textwidth]{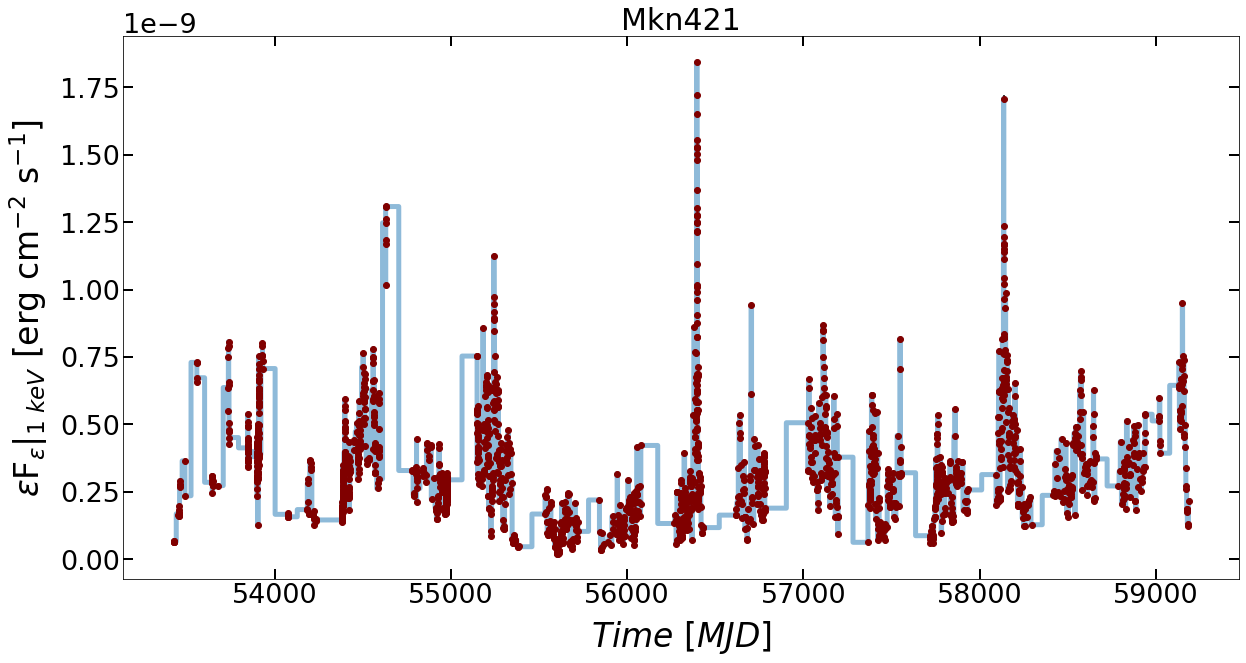}
\hfill
\includegraphics[width=0.47\textwidth]{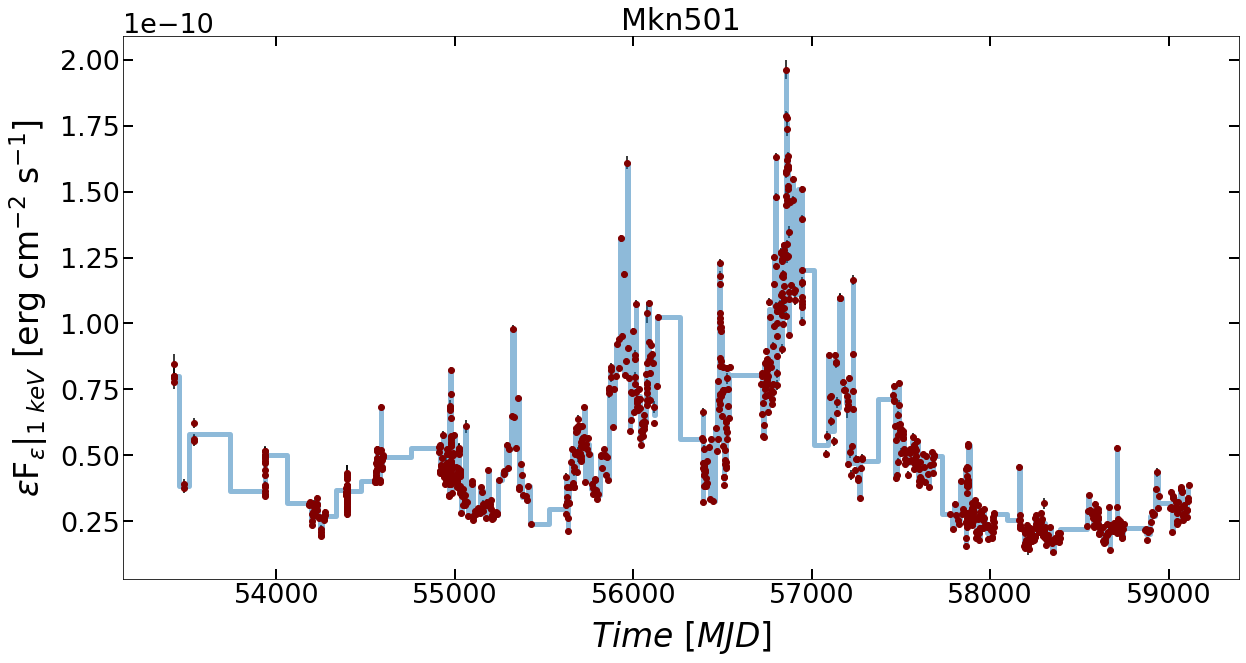} \\
\vspace{0.2in}
\includegraphics[width=0.47\textwidth]{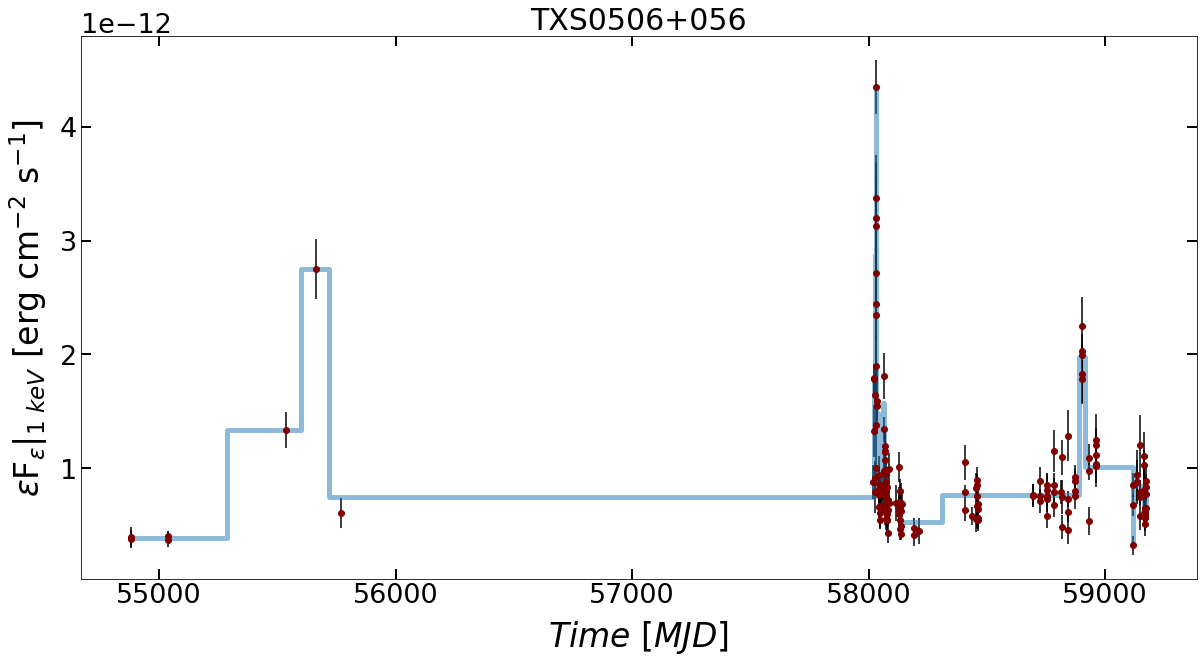}
\hfill
\includegraphics[width=0.47\textwidth]{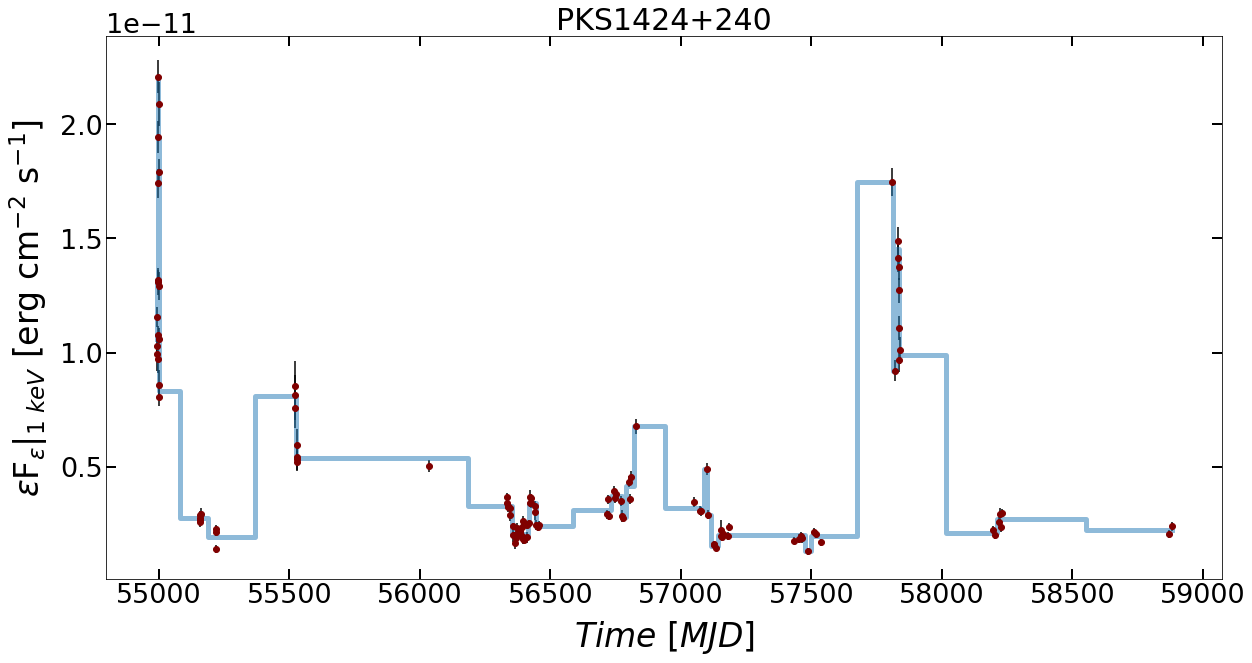}  \\
\vspace{0.2in}
\includegraphics[width=0.47\textwidth]{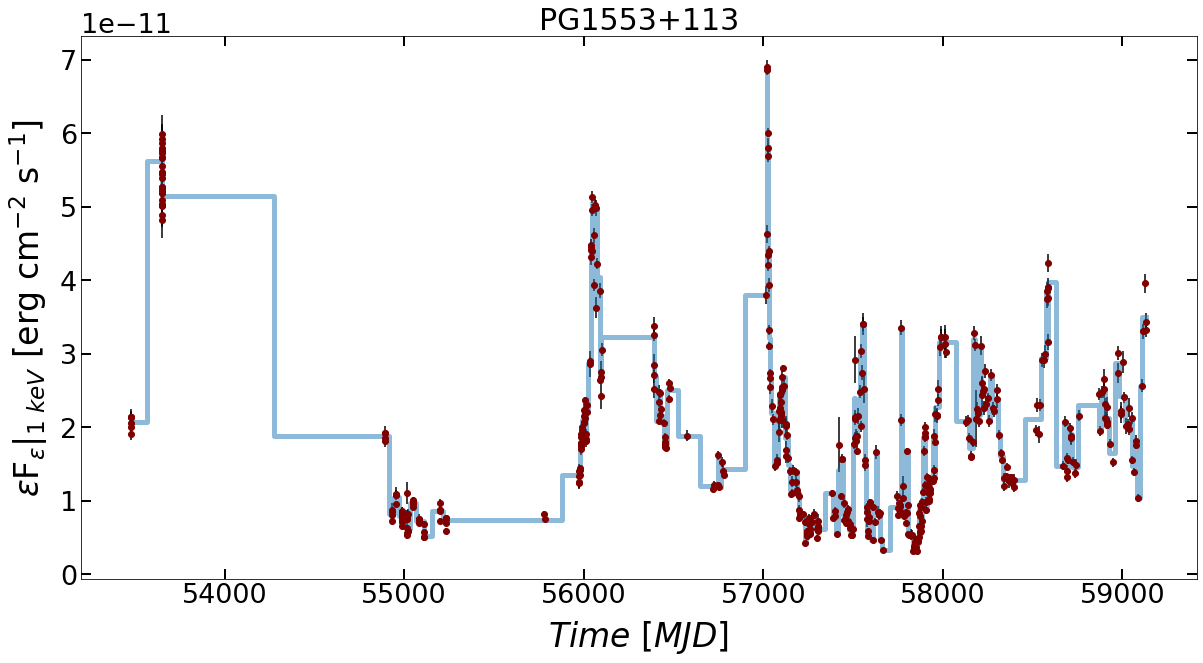}
\hfill
\includegraphics[width=0.47\textwidth]{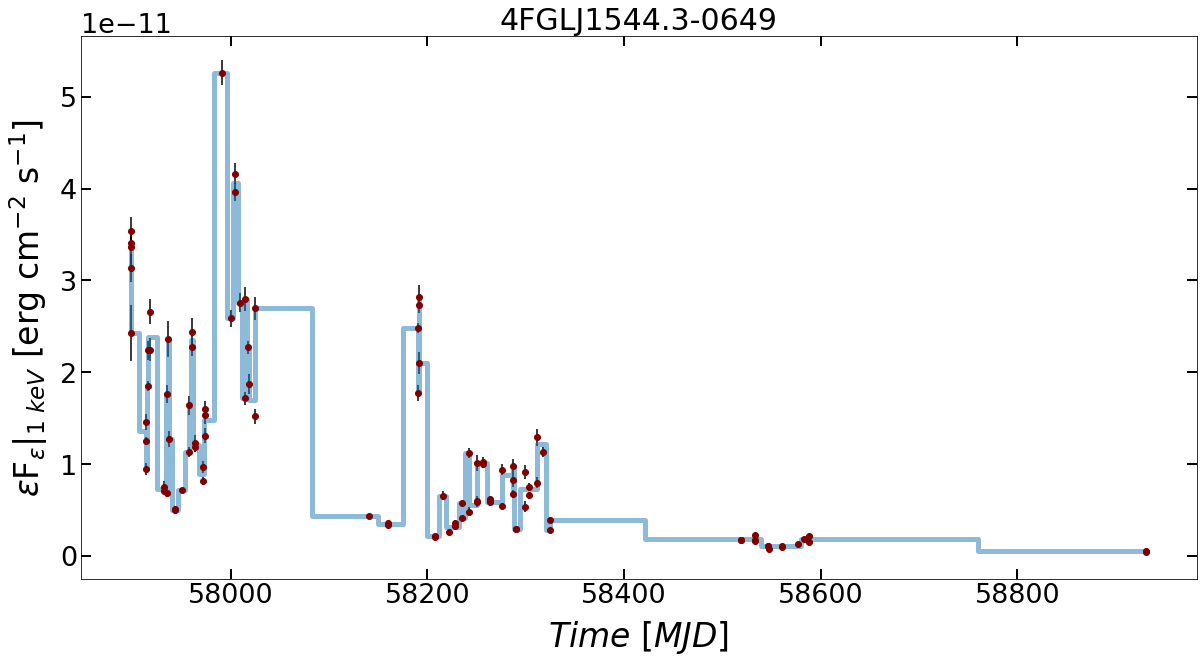} \\
\vspace{0.2in} 
\includegraphics[width=0.47\textwidth]{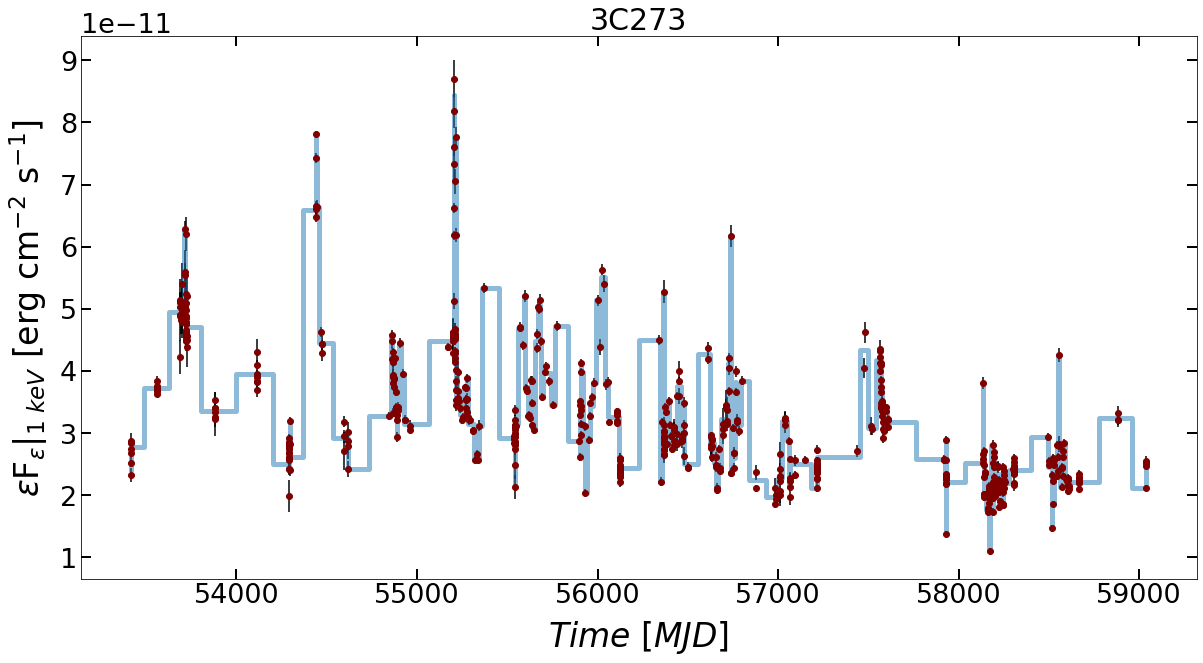}
\hfill
\includegraphics[width=0.47\textwidth]{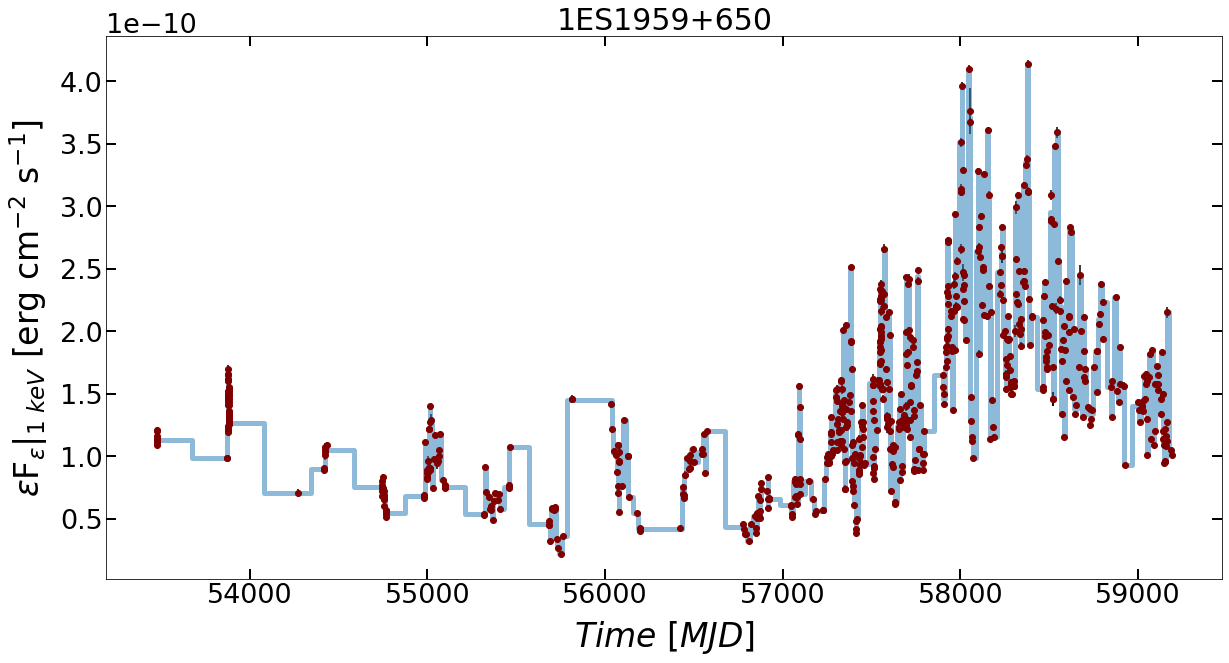}
\caption{1~keV light curves of blazars from our sample (symbols). Error bars indicate the 68\% uncertainty in flux. Solid lines show the Bayesian block representation of the light curves. Long horizontal lines with no sampling between a data point and a new block do not guarantee a stable flux. }
\label{fig:lc}
\end{figure*}

\subsection{Bayesian blocks and definition of flares}\label{sec:flares}
To search for X-ray variability we apply a Bayesian blocks algorithm to every X-ray light curve of our sample\footnote{All light curves with the Bayesian block representation can be found at \url{https://stamstath.wixsite.com/1kevxrtlc}}. The algorithm finds the optimal segmentation of the data taking into account the statistical fluctuations from the measurement errors. This allows us to represent each light curve by a series of contiguous ``blocks'' where the flux is considered to be constant. This  block representation provides an objective way to detect significant variations in a light curve regardless of variations in gaps or exposure.
We note however that we cannot probe variations in flux shorter than the typical duration of an XRT snapshot, as this is the building block of our light curves.

We use the {\tt astropy} implementation of the Bayesian Blocks algorithm \citep{Astropy:2018} described in \cite{Scargle:2013}, with the option of ``measures'' in the fitness function and false alarm probability $p_{0} = 0.1$. This parameter is related to the prior on the number of bins, ncp$_{\rm prior}$, and the actual number of data points $N$ as ncp$_{\rm prior}=4 - \ln(73.53 \,   p_0  \, N^{-0.478})$.
While $p_0$ affects the total number of blocks building the light curve, we expect no big differences in the derived flaring states and total number of neutrino events for $p_0 \sim 0.01-0.1$ (for details, see Appendix~\ref{app:p0}). 

The Bayesian block representation of the light curves  presented in Fig.~\ref{fig:lc} is indicated by solid lines. 
The height of each
block is the statistical mean of all  flux measurements belonging to it. 
 Large gaps between consecutive data points are represented by blocks with long duration. These long horizontal lines have usually no sampling between a data point and a new block. So, interpretation of these blocks as periods of stable flux should be made with caution. We will discuss in more detail the impact of long-duration blocks on our results later in Sec.~\ref{sec:flares-results}.

Several definitions of flares have been proposed in the literature~\citep[e.g.][]{2009A&A...502..499R,  2016A&A...593A..91A,  2019ApJ...877...39M}. Flares could be, for instance, defined by an increase in the block flux by at least a factor of 2. In this case, a flare could be comprised of several rising blocks in a row. Alternatively, flares could be defined using the light curves directly and not their Bayesian block representation. For instance, \cite{2013MNRAS.430.1324N} defined flares as periods of time containing a local maximum in flux during which the flux exceeds half of the peak value. This definition would not allow any two flares to overlap. Flares could also be identified by finding the local maxima of a light curve, and then be fitted using pre-defined functional forms \citep[e.g., a piece-wise exponential functions][]{1999ApJS..120...95V, 2010ApJ...722..520A, 2018ApJ...856...95A}. An alternative way of studying flux variability in blazar light curves and studying the properties of flares was presented by \cite{2018ApJ...866..137L}. These authors used used a Bayesian hierarchical model that treats each light curve as a superposition of flares with different shapes. In this approach, a peak in the light curve could be composed by several overlapping `flares'. While the definition of flares may affect the statistical properties of the inferred flaring states (i.e. duration and fluxes), it is not expected to affect significantly the fluence, hence the expected number of neutrino events from flares. 
In this work we use the following definitions.

\noindent  {\bf Definition 1} (Flare).
Flare is any block with flux $f_{\rm B}$ exceeding the mean value $\mu$ of all flux measurements by a factor of $n\sigma$. Here, $n$ is an integer and $\sigma$ is the standard deviation of the flux  measurements. \par 
\noindent 
{\bf Definition 2} (Flare duration).
The duration of a flaring block, $\Delta t$, is used as a proxy of the duration of the X-ray flare, which is needed for the calculation of the neutrino events (see equation~\ref{eq:events}). 
% We return to this point later in Section~\ref{sec:neutrinos-results}.
% \begin{definition}[Flare]
% Flare is any block with flux $f_{\rm B}$ exceeding the mean value $\mu$ of all flux measurements by a factor of $n\sigma$. Here, $n$ is an integer and $\sigma$ is the standard deviation of the flux  measurements. 
% \end{definition}
% \begin{definition}[Flare duration]
% The duration of a flaring block, $\Delta t$, is used as a proxy of the duration of the X-ray flare, which is needed for the calculation of the neutrino events (see equation~\ref{eq:events}). 
% We return to this point later in Section~\ref{sec:neutrinos-results}.
% \end{definition}

Therefore, when two or more consecutive blocks are found to overcome the flare threshold they are treated as separate flares.
Depending on $n \ge 1$ there is a probability that the  selected flaring block is a true enhancement in the photon flux of the source or a fluctuation of the average flux level. Wanting to investigate a likely relation between the flaring block flux and duration, we also distinguish flares in two types as follows
\begin{itemize}
    \item Type A: $\mu+\sigma < f_{\rm B} < \mu+3\sigma$  
    \item Type B: $f_{\rm B} > \mu+3\sigma $  
\end{itemize}
This classification may be phenomenological but it can help us investigate if a certain type of flares has a larger contribution to the neutrino fluence of a source (i.e. we expect higher neutrino flux from a Type B flare of the same duration than a Type A flare for a given source).
\begin{figure}
    \centering
    \includegraphics[width=0.47\textwidth]{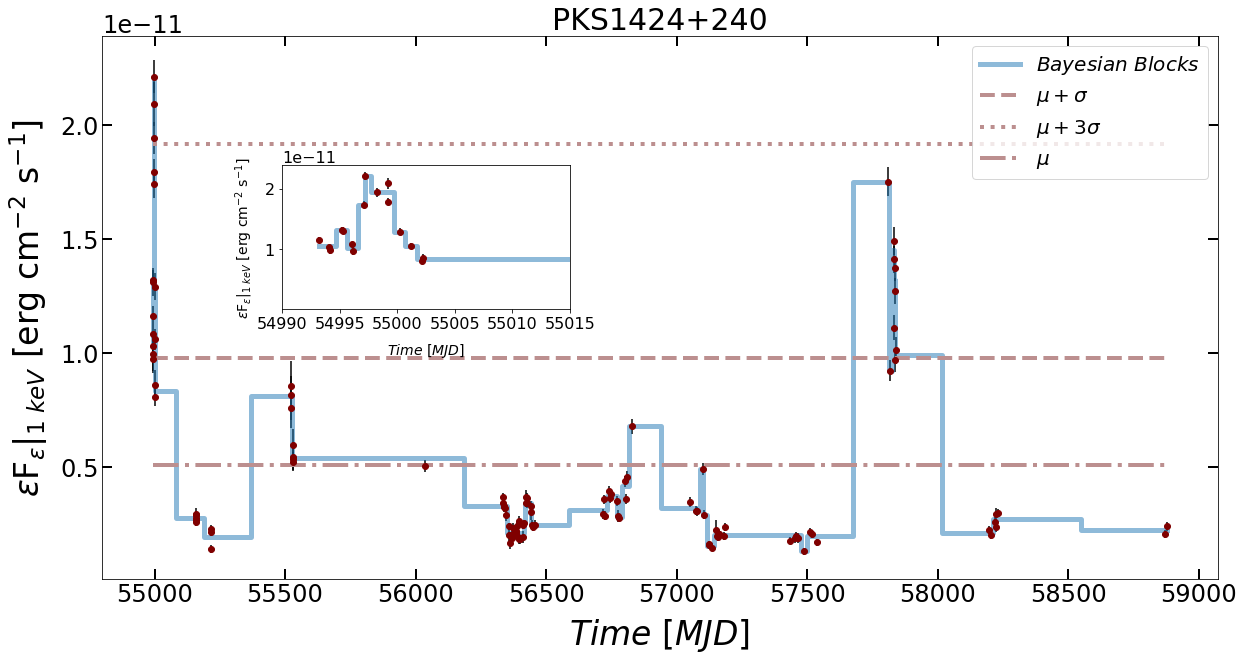}
    \caption{1~keV light curve of PKS~1424+240 with the two flux threshold values used to classify flares indicated by horizontal lines (see text for details). The inset plot shows a zoom in to the early-time portion of the light curve. }
    \label{fig:lc-flares}
\end{figure}

The identification of flares\footnote{Henceforth, we use the terms `flares' and `flaring blocks' interchangeably.} is exemplified in Fig.~\ref{fig:lc-flares}, where we show the full 1~keV light curve of PKS~1424+240 and the two flux thresholds discussed above (solid and dashed line) and the dashed-doted line denotes the mean of all flux measurements. At early times ($\sim 54990-55005$~MJD), the source was in a flaring state. If we zoom into that portion of the light curve (see inset plot), we can identify several blocks with short durations ($\sim 0.6$~d) and $\mu+\sigma < f_{\rm B} < \mu+3\sigma$ (Type A), while only 2 blocks exceed the $\mu+3\sigma$ threshold (Type B). 
The duration and flux distributions of all flares identified in the sample are presented in Section~\ref{sec:flares-results}.

\section{Expected neutrino event counts}\label{sec:neutrino-counts}
The expected number of muon plus antimuon neutrinos from an X-ray flare can be calculated as  
\begin{equation}
\mathcal{N}_{\nu_{\mu}+\bar{\nu}_{\mu}}=\frac{1}{3}\int_{t_{\rm ini}}^{t_{\rm end}}{\rm d}t \int_{E_{\nu, \min}}^{E_{\nu, \max}}{\rm d}\varepsilon_{\nu} \, A_{\rm eff}(\varepsilon_{\nu},\delta) \frac{F_{\nu+\bar{\nu}}(\varepsilon_{\nu},t)}{\varepsilon_{\nu}},
\label{eq:events}
\end{equation}
where we assumed vacuum neutrino mixing and used 1/3 to convert the all-flavour to muon neutrino flux. Moreover, $t_{\rm ini}$ and $t_{\rm end}$ define the duration of the X-ray flare as $\Delta t = t_{\rm end}-t_{\rm ini}$ and $A_{\rm eff}(\varepsilon_{\nu},\delta)$ is the energy-dependent and declination-dependent point-source effective area of IceCube \citep{Aartsen_2020, 2021arXiv210109836I}. \swift \, observations for certain sources, such as Mkn~421, are available since 2005, well before the starting date of IceCube operations. We therefore use different effective areas for our calculations (see Table~\ref{tab:Aeff}) depending on the configuration of IceCube at the time of the flare. For this purpose, we check if the midpoint, $(t_{\rm ini}+t_{\rm end})/2$, of a flare block falls in a specific season of IceCube operation and adopt the corresponding effective area. For flares occurring before the start of IC40, we set the number of events equal to zero. For the integration over energies we set $E_{\nu, \min}=100$~TeV and use the maximum energy to which $A_{\rm eff}$ is computed as $E_{\nu, \max}$.

\begin{table}
    \centering
    \begin{tabular}{c|c}
    \hline 
    IceCube configuration     &  Season (MJD) \\
    \hline
    IC40     &  $54562 - 54971$ \\
    IC59     & $54971 - 55347$ \\
    IC79     & $55347 - 55694$ \\
    IC86-I   & $55694 - 56062$ \\
    IC86-II  & $>56062$ \\ 
    \hline 
    \end{tabular}
    \caption{Point-source effective areas for different configurations used in our analysis. Data are taken from  \citealt{Aartsen_2020, 2021arXiv210109836I}.}
    \label{tab:Aeff}
\end{table}

The neutrino energy flux, $F_{\nu+\bar{\nu}}(\varepsilon_{\nu},t)$, is computed using equations (\ref{eq:Fnu})-(\ref{eq:fX}). To account for a non-hadronic origin of the non-flaring X-ray emission, as illustrated in Fig.~\ref{fig:sed}, we subtract from all X-ray flux measurements, $F_{\rm X}$,  the mean of the 0.5-10 keV energy fluxes. We discuss how this choice affects our neutrino predictions in Section~\ref{sec:discussion}. 
 
Depending on the number of flux measurements contained within a flare block with duration $\Delta t$, we treat the time integral of equation~(\ref{eq:events}) differently. More specifically, if there are multiple flux measurements within the block of the flaring state  (i.e. $N>1$), then the predicted muon and antimuon number of neutrinos is estimated as 

\begin{equation}
\mathcal{N}_{\nu_{\mu}+\bar{\nu}_{\mu}}\approx\frac{1}{3}\sum_{i=1}^{N-1}\Delta t_i \frac{F_{0,i}\mathcal{I}_i+ F_{0,i+1}\mathcal{I}_{i+1}}{2} + \langle F_0 \mathcal{I}\rangle \left(\Delta t - t_{\rm N} + t_{1}\right),
\label{eq:events1}
\end{equation}
where 
the index $i$ runs over the number of flux measurements, $F_{0,i} \equiv F_0 (t_i)$,  $\Delta t_i \equiv t_{i+1}- t_i$, $\langle ... \rangle$ denotes the mean over the flux measurements, and 
\begin{equation} 
\mathcal{I}_i \equiv  \int_{E_{\nu,\max}}^{E_{\nu,\min}} \! \! {\rm d}\varepsilon_{\nu} \, A_{\rm eff}(\varepsilon_{\nu},\delta)  \varepsilon_{\nu}^{-s-1} e^{-\varepsilon_{\nu}/\varepsilon_{\nu, \rm c}(t_i)}.
\label{integral1}
\end{equation}
The second term on the right-hand side of equation~(\ref{eq:events1}) takes into account the contribution from the block outside the time window of flux measurements ($t>t_N$ and $t< t_1$). The peak neutrino energy $\varepsilon_{\nu, \rm c}$ is given by equation~(\ref{eq:Enu}) after replacing $\varepsilon_{\rm keV}$ with the peak energy of the X-ray spectrum in $\varepsilon F_{\varepsilon}$ space ($\varepsilon_{\rm pk}$). Depending on the photon index $\Gamma$ of the best-fit power-law spectrum in the 0.5-10 keV energy range, which can vary between measurements, we consider two options. If $\Gamma <2 \,  (>2)$, then $\varepsilon_{\rm pk} = 10 \, (0.5)$~keV, and if $\Gamma=2$ we set $\varepsilon_{\rm pk}=2.23$~keV (i.e. the logarithmic mean of the energy band). While it could be possible that the true peak energy of the X-ray spectrum (in $\varepsilon F_{\varepsilon}$) might lie outside the 0.5-10 keV range we prefer not to extrapolate but rely instead only on narrow-band spectral information. 

If there is only one flux measurement within the block of a flaring state, we cannot do much better than to assume that the 0.5-10~keV energy flux and the peak neutrino energy remain constant over the time window of the flare.  In this case, equation~(\ref{eq:events}) simplifies into the following one
\begin{equation}
\mathcal{N}_{\nu_{\mu}+\bar{\nu}_{\mu}}=\frac{\Delta t F_{\rm 0.5-10~keV}}{2.7~\varepsilon_{\nu, \rm c}^{-s+1} } \mathcal{I}, 
\label{eq:events2}
\end{equation}
where $F_{\rm 0.5-10~keV}$ is the integrated mean-subtracted flux between 0.5 keV and 10 keV energies, and $\Delta t$ is the duration of each flaring state extracted from the Bayesian blocks analysis of the 1~keV light curve.

\begin{figure}
\includegraphics[width=0.47\textwidth]{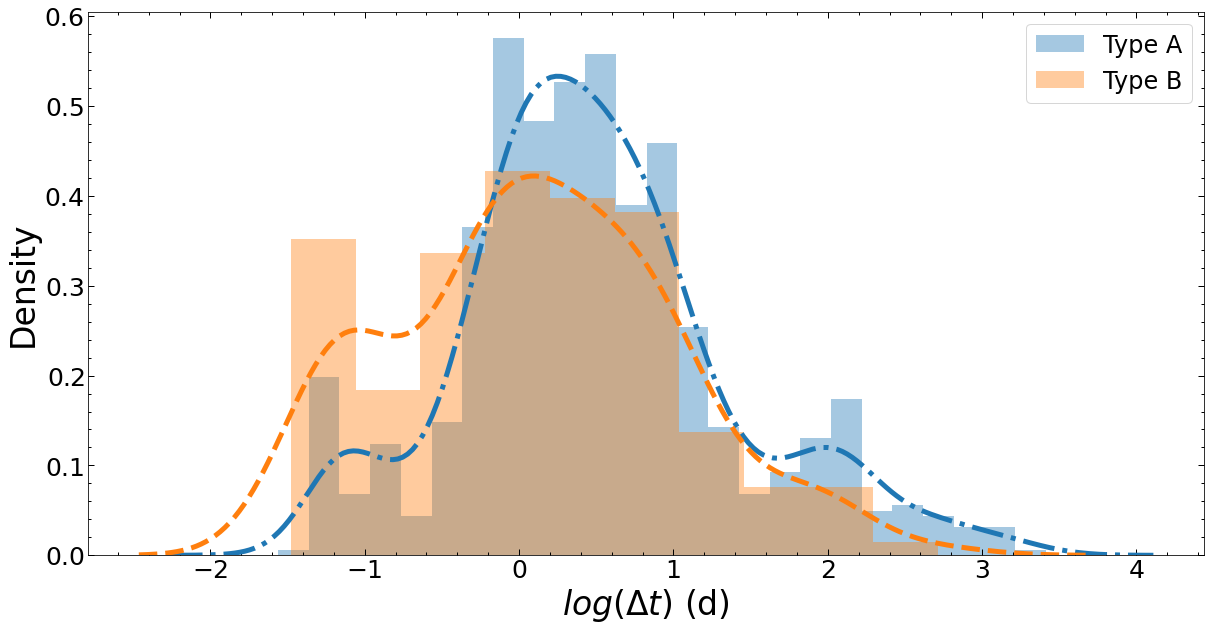} 
\includegraphics[width=0.47\textwidth]{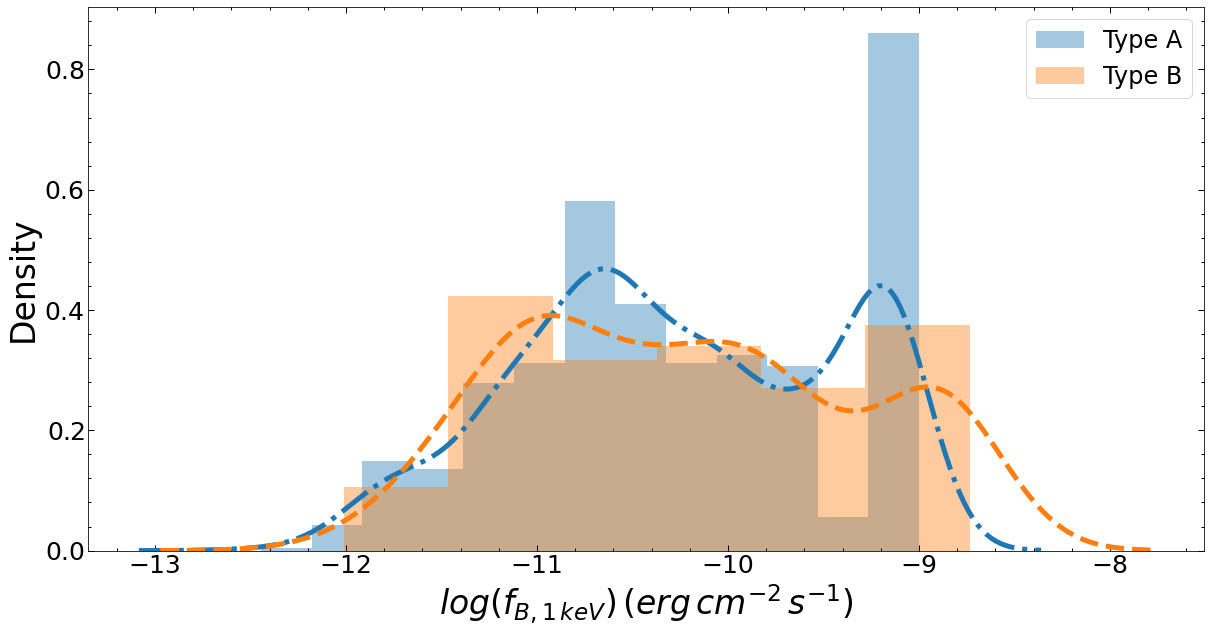}
\caption{Normalized distributions of durations ($\Delta t$) and fluxes at 1~keV ($f_{\rm B, 1~{\rm keV}}$) of blocks classified as flares of Types A and B (coloured histograms). Data are binned using the Freedman-Diaconis estimator, hence the differences in bin size. Dash-dotted and dashed lines show the representation of the data for flares of Type A and B respectively using a continuous probability density curve.}
\label{fig:1Dhisto}
\end{figure}   

The conventional muon plus antimuon neutrino atmospheric flux on the surface of the Earth forms a background at high energies for searches of point-like neutrino sources. In our model, the source neutrino spectrum typically peaks at $\varepsilon_\nu\gtrsim 1$~PeV and the neutrino number of events is computed above $100$~TeV. 
Above this energy the contribution of the atmospheric background (declination-averaged) is $\sim 0.0007$ events per year and can be safely neglected in most cases \citep[see also][for Mkn~421]{Petropoulou_2016}. For completeness, we compute the yearly rate of atmospheric muon and antimuon neutrinos above 100 TeV coming from the direction of each source (see last column of  Table \ref{tab:sample}). We approximate the conventional muon plus antimuon neutrino atmospheric flux by a  power law with index $\sim -3.7$ ~\citep{2007PhRvD..75d3006H}, and  treat this component as purely isotropic.
For the normalization at 100 TeV of the atmospheric muon and antimuon neutrino fluxes averaged over the zenith angle we use the mean value of the model predictions as presented in fig.~33 of  \cite{article}. The expected muon and antimuon number from the atmospheric background is then calculated using equation~(\ref{eq:events}) by integrating over energy, time, and solid angle assuming that the neutrino flux and effective area (we use IC86-II configuration) are constant. We integrate over a typical angular resolution of 1 degree to estimate the expected neutrino number. Thus, the integral over the solid angle in equation~(\ref{eq:events}) reduces to a constant.  

\section{Results}\label{sec:results}
\subsection{X-ray flares}\label{sec:flares-results}
Using the 1~keV X-ray light curves we find in total 967 flaring states (of both types). About 22\%  of flaring states ($217 / 967$) are attributed to Mkn~421, which is one of the brightest and, as a result, best monitored blazars at all wavelengths.

Fig.~\ref{fig:1Dhisto} presents the normalized distributions of durations ($\Delta t$) and fluxes at 1~keV ($f_{\rm B, 1~{\rm keV}}$) of blocks classified as flares. Histograms of different flare types are displayed with different colours. The contribution of Mkn~421 to the flaring sample is evident by the highest flux bin in the histogram of flare fluxes (of both types). Instead of testing whether the two groups of flares are different, we derive an estimate of how different their mean values and standard deviations are using a Bayesian estimation tool \citep{Kruschke2013} implemented in PyMC3\footnote{\url{https://docs.pymc.io/notebooks/BEST.html}}. For the difference of means in flux (duration), at least 99\% of the posterior probability values are less (greater) than zero. This suggests that the group means are credibly different. The differences in the standard deviations of flux and duration are, however, smaller. These results do not necessarily reflect intrinsic differences between flare types, as they could arise from observational biases related to the irregular sampling of XRT observations. For instance, states with higher fluxes are more likely to be observed multiple consecutive
times, while low-flux states are less frequently observed (see also Fig.~\ref{fig:lc}).

\begin{figure}
    \centering
    \includegraphics[width=0.47\textwidth]{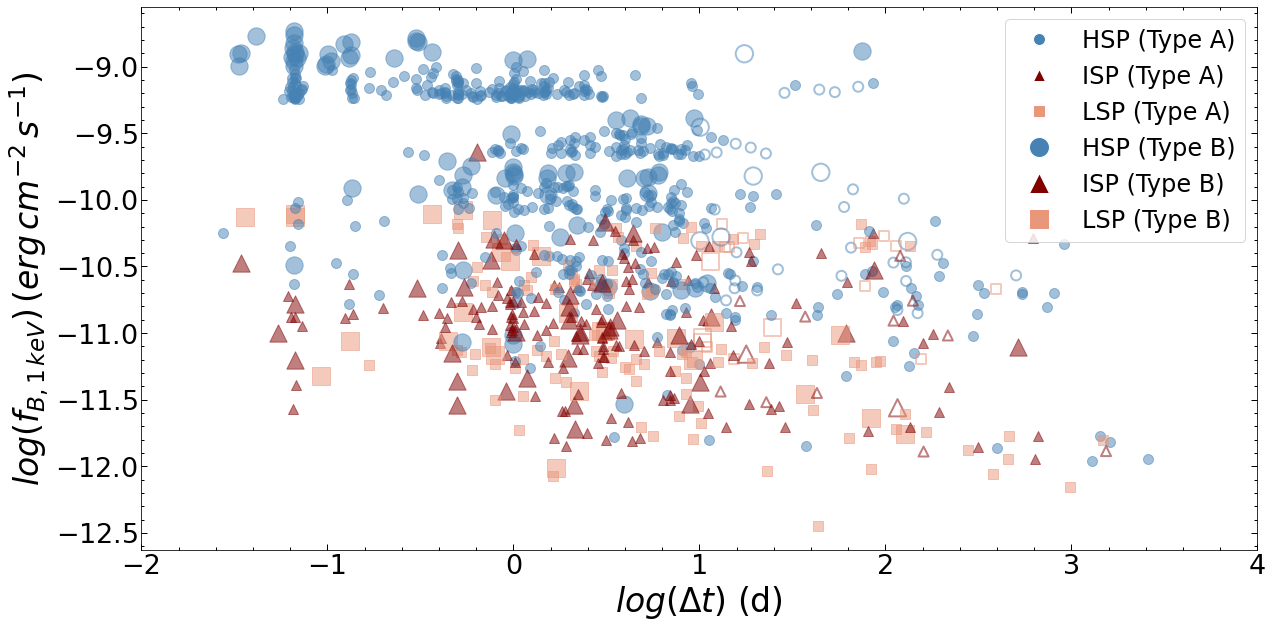}
    \caption{Block fluxes at 1~keV versus duration of blocks classified as flares of Types A and B. Different coloured symbols indicate the spectral blazar class (see inset legend). Open symbols indicate blocks with $\Delta t>10$~d and only one flux measurement within this interval.
    }
    \label{fig:flux-duration}
\end{figure}

Fig.~\ref{fig:flux-duration} summarizes our findings by showing the block flux computed from the 1~keV light curves, $f_{\rm B, 1~{\rm keV}}$, as a function of the block duration $\Delta t$. Different symbols indicate blazars of different spectral types, namely HSP (circles), ISP (triangles), and LSP (squares). Flares from HSP sources are on average brighter than those produced by ISP or LSP objects. The clump of points with $f_{\rm B}\gtrsim 3\times10^{-10}$~erg cm$^{-2}$ s$^{-1}$ corresponds to Mkn~421, which dominates our sample both in terms of flare number and flare brightness. We find no clear evidence for a linear relation between the flux and duration of blocks identified as flares or of the type of flares with either duration or flux. A  careful statistical analysis of the flare properties is unwarranted at this point because of observational biases affecting our sample. For instance, a comparison of the flare fluxes, durations, and duty cycles between sources with very different X-ray coverage (e.g. Mkn~421 and TXS~0506-056) would not yield meaningful results. We will return to this point in Section~\ref{sec:discussion}.

Open symbols in Fig.~\ref{fig:flux-duration} indicate blocks with $\Delta t > 10$~d containing only one flux measurement (for the neutrino expectation from such flares, see Section~\ref{sec:neutrino-counts}). In fact, after visual inspection of the Bayesian block representation of all light curves, we find that most blocks with $\Delta t \gtrsim 60$~d contain $\sim1-2$ XRT snapshots (see e.g. second and third blocks from the start of the light curve of TXS~0506+056 in Fig.~\ref{fig:lc}). Using the flux of a couple XRT snapshots with total duration of a few ks as a proxy for the source flux state on week-long or even month-long periods introduces big uncertainties in the predicted neutrino fluence. Hence, if the block duration is $>10$~d and contains only one XRT observation, we set $\Delta t =1$~d equation~(\ref{eq:events2}), which is close to the most probable value of the duration distribution (see Fig.~\ref{fig:1Dhisto}). Similarly, most blocks with $\Delta t \sim 30-60$~d contain a handful of measurements clustered in time, occupying only a small fraction of the total block duration. Such month-long blocks are a result of large gaps between \swift \, observations (see e.g. the light curve of PG~1553+113 in Fig.~\ref{fig:lc}) caused by the lack of all-sky X-ray monitoring. Because we cannot predict the behaviour of the source during these long periods, we will also report the expected number of neutrinos from each source after excluding these blocks (for details, see Section~\ref{sec:neutrinos-results}).

\subsection{Neutrinos from X-ray flares}\label{sec:neutrinos-results}
\begin{figure}
\includegraphics[width=0.47\textwidth]{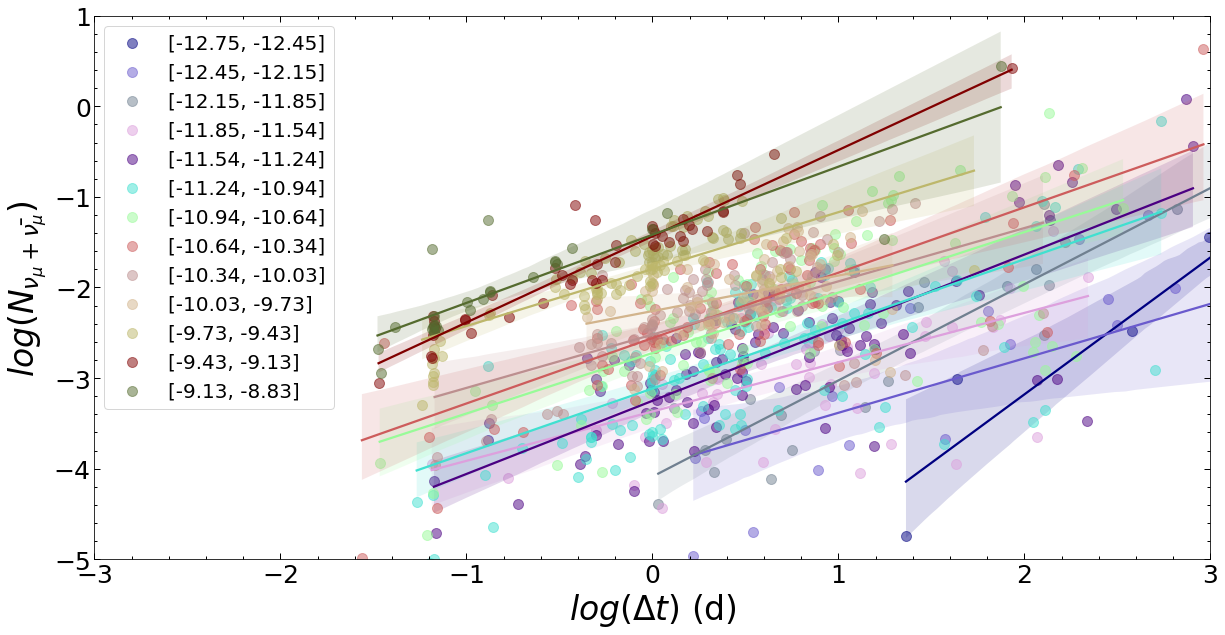}
\caption{Expected muon and antimuon neutrino number from X-ray flares versus the duration of the flare as defined by the Bayesian block algorithm (in logarithmic scale). Symbols are colour coded according to the 1~keV flux of the flares (see also Fig.~\ref{fig:flux-duration}). Solid lines show linear regression model fits to the data, and shaded regions indicate the 95\% confidence intervals.}
\label{fig:eve_dur}
\end{figure}

Fig.~\ref{fig:eve_dur} shows the predicted number of muon and antimuon neutrino events from X-ray flares occurring after 54562~MJD as a function of the block duration. Different colours are used to indicate the 1~keV flare flux (see inset legend). We used the same flux bins as those determined by the Freedman-Diaconis estimator for the flux histogram shown in Fig.~\ref{fig:1Dhisto} (bottom panel). For fixed duration, flares with higher X-ray fluxes are found to produce a higher number of events   compared to flares with lower X-ray fluxes. This finding basically reflects the model's main assumption, namely $F_{\nu+\bar{\nu}}\approx F_{\rm X}$ (see also Section~\ref{sec:model}). Each intermediate flux state will fall inside these boundary lines. The relation between the duration of each flaring state and the predicted number of events (in logarithmic space) is well described by a linear function, as shown by the linear regression fit to the data (see solid lines). The correlation of $\mathcal{N}_{\nu_{\mu}+\bar{\nu}_{\mu}}$ with $\Delta t$ is another demonstration of the lack of strong correlation between the X-ray flux and duration of flares (see also Fig.~\ref{fig:flux-duration}). Thus, flares with similar flux will produce more neutrinos if they last longer. The scatter of the neutrino number within one flux bin is mostly a result of the declination-dependent area of the detector (see equation~(\ref{eq:events})). For instance, the scatter is significantly reduced in the two bins with the highest X-ray fluxes ($-12.75<\log(f_{\rm B, 1~{\rm keV}})<-12.15$) that are dominated by one source (Mkn~421). 

\begin{figure}
\includegraphics[width=0.47\textwidth]{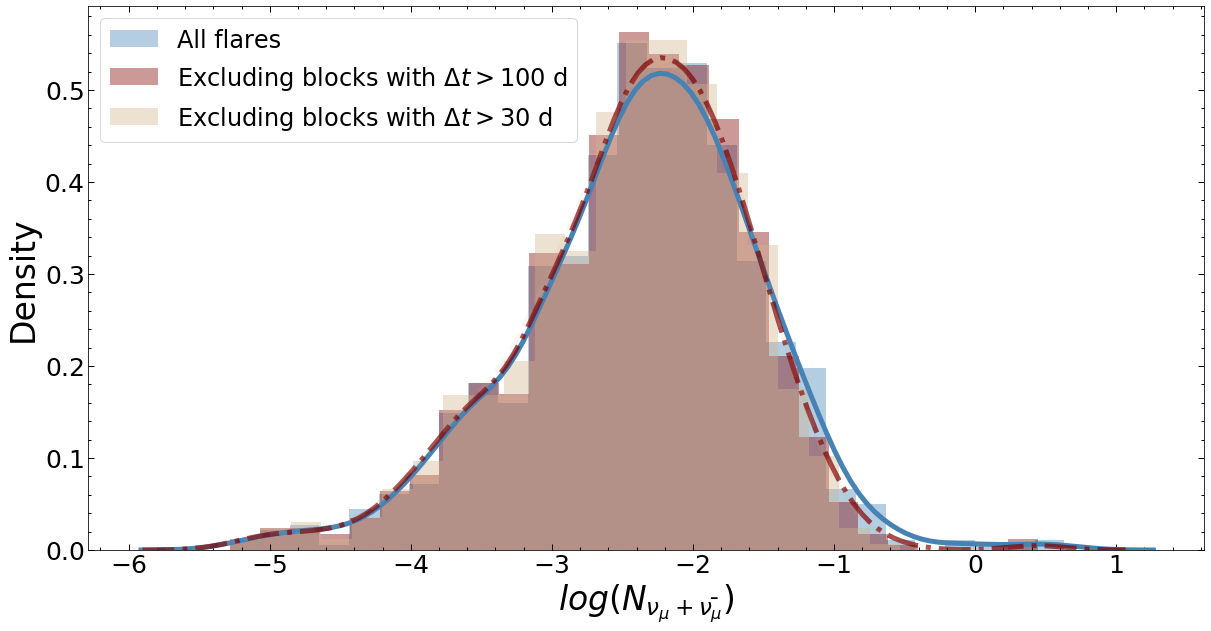}
\caption{Normalized distribution of the expected muon and antimuon neutrino number from X-ray flares (in logarithmic scale). Distributions obtained after removing blocks with $\Delta t > 100$~d and 30~d are overplotted for comparison.}
\label{fig:1Dhisto-neutrinos}
\end{figure}

\begin{figure}
\includegraphics[width=0.47\textwidth]{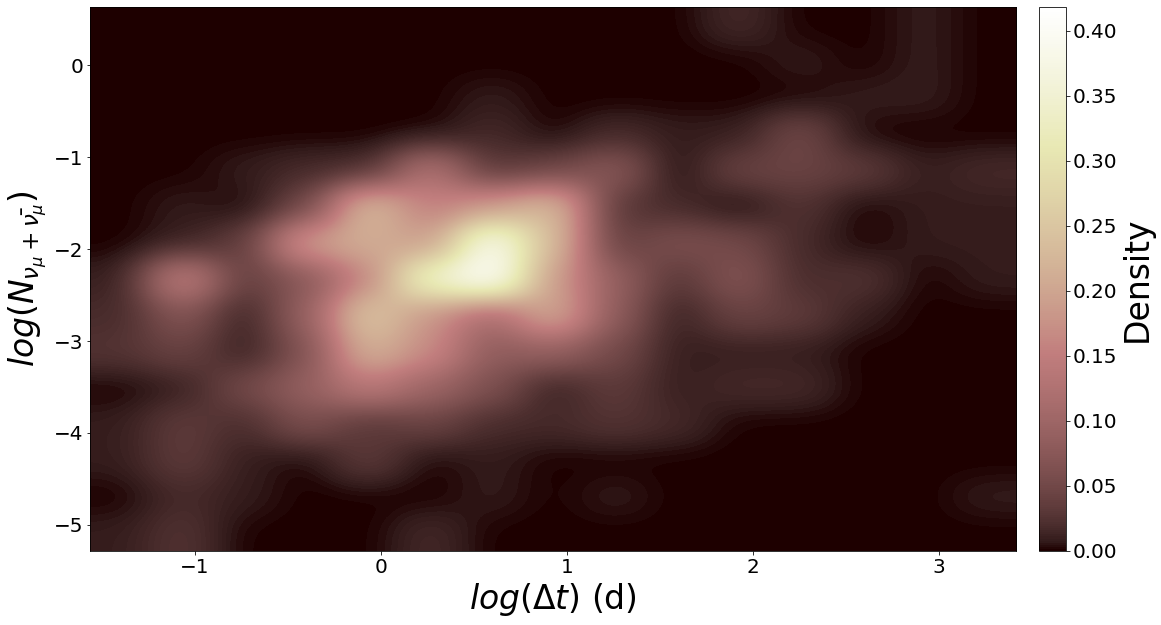}
\caption{Two-dimensional histogram (in logarithmic scale) of the number of muon and antimuon neutrinos expected from flares of duration $\Delta t$. The histogram is normalized so that the area underneath it integrates to 1.}
\label{fig:2Dhisto}
\end{figure}

Fig.~\ref{fig:1Dhisto-neutrinos} shows the distribution of the predicted number of muon and antimuon neutrino events from all flares happening after 54562~MJD (blue), which has a median of $\simeq 0.01$~events. We also plot the histograms of $\log( \mathcal{N}_{\nu_{\mu}+\bar{\nu}_{\mu}})$ after excluding the contributions of blocks with $\Delta t > 100$~d (maroon) and 30~d (tan). While the choice of the specific time windows is not strict, it is motivated by the following:  (a) blocks with $\Delta t > 60$~d, in a plethora of cases, contain  XRT snapshots that have time separations similar to the duration of the block itself; (b) after visual inspection of the Bayesian representation of the light curves in our sample, we find that blocks with $\Delta t >30$~d  often contain clustered XRT measurements that occupy only a small fraction of the block duration. As discussed in Section~\ref{sec:flares-results}, such month-long blocks are usually a result of large gaps between \swift \, observations (see e.g. the light curve of PG~1553+113 in Fig.~\ref{fig:lc}). Neutrino fluences computed by assuming a constant flux level for such a long time are therefore highly uncertain. Nonetheless, Fig.~\ref{fig:1Dhisto-neutrinos} shows that the general shape, including the mean and median, of the event distribution does not change after removal of blocks with $\Delta t>30$~d. Indeed, there are only a few blazars in our sample whose main contribution to the neutrino number comes from flares with $\Delta t>30$~d. These findings suggest that the bulk of the neutrino events of our sample originates from flares with much shorter durations whose neutrino fluence predictions are more robust. To better illustrate this, we present the two-dimensional density map of $\log(\mathcal{N}_{\nu_{\mu}+\bar{\nu}_{\mu}})$ versus $\log(\Delta t)$ in  Fig.~\ref{fig:2Dhisto}. Indeed, the highest density is observed for $\Delta t \sim 1-10$~d and $\mathcal{N}_{\nu_{\mu}+\bar{\nu}_{\mu}} \sim 0.01$. Hence, blocks with $\Delta t \gtrsim 30$~d that may be sources of large systematic uncertainties in the neutrino fluence do not seem to affect the neutrino expectation of the whole sample.

\begin{figure*}
\includegraphics[width=0.95\textwidth]{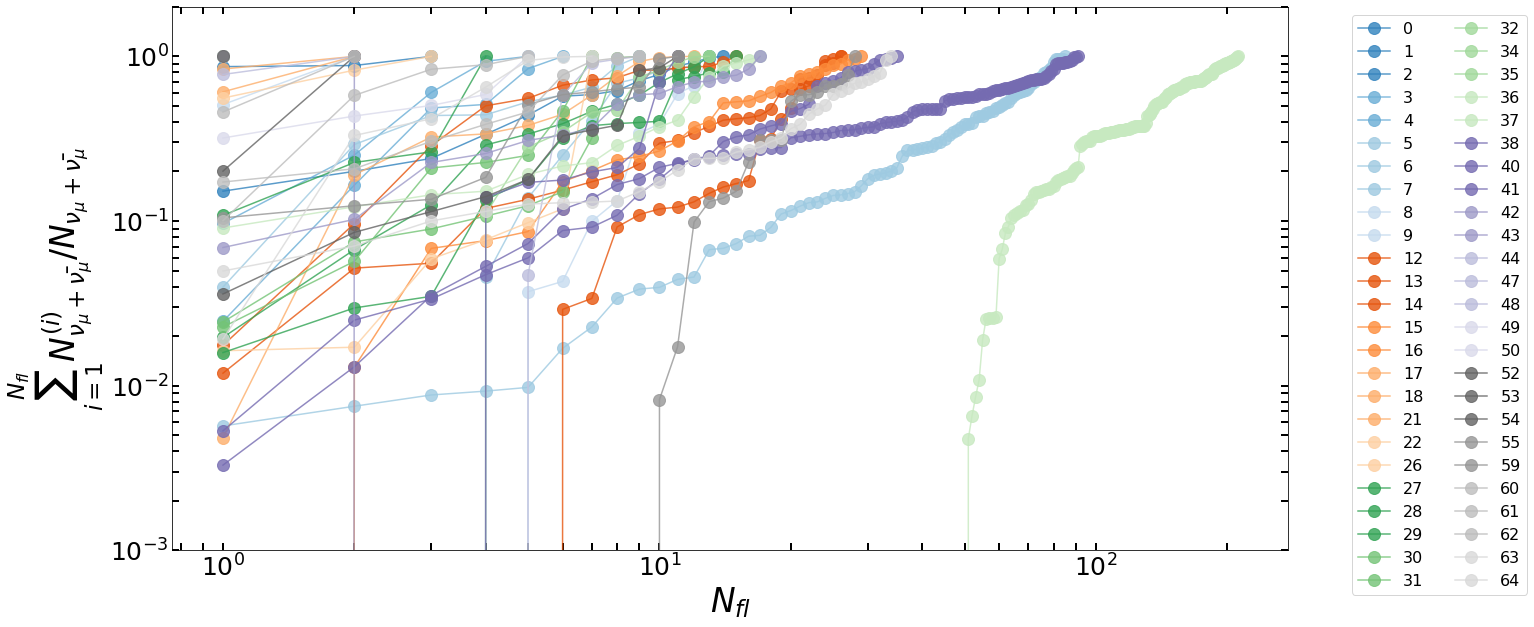}
\caption{Cumulative number of muon and antimuon neutrino events expected from individual sources and $N_{\rm fl}$ X-ray flares with $\Delta t <30$~d. Coloured symbols correspond to different sources, as indicated by their indices in the inset legend (for the source names see Table~\ref{tab:sample}).}
\label{fig:cumulative}
\end{figure*}

\begin{figure*}
\includegraphics[width=0.95\textwidth]{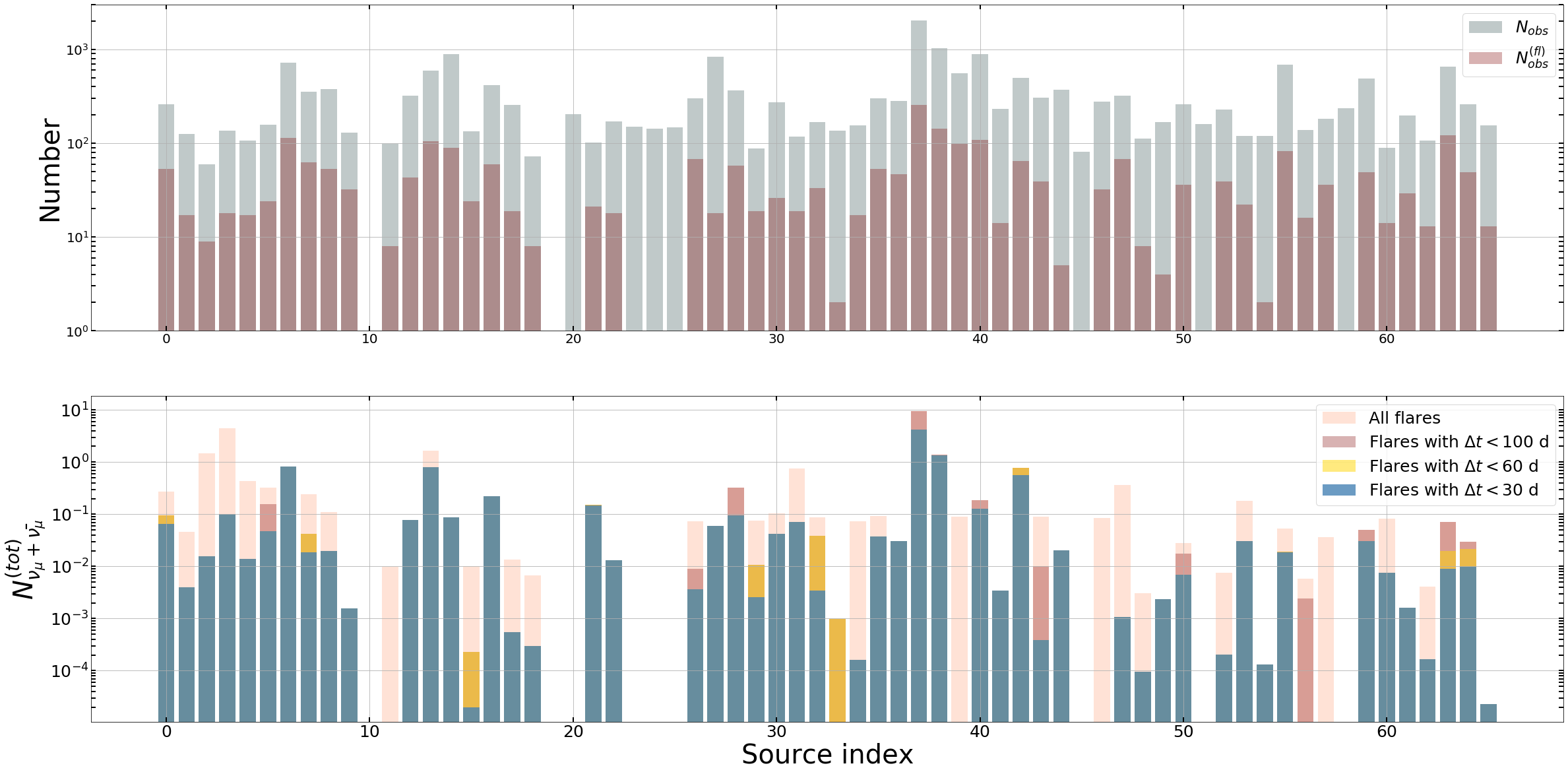}
\caption{\textit{Bottom panel:} Stacked number of  muon and anti-muon neutrinos expected from  X-ray flares of individual sources (coloured bars). Different colours indicate the expected number when different cuts on the block duration are made (see inset legend). \textit{Top panel:} Number of XRT observations ($N_{\rm obs}$) and number of XRT observations belonging to blocks identified as flares ($N^{(\rm fl)}_{\rm obs}$) per source. All values are listed in Table~\ref{tab:sample}.}
\label{fig:barplot}
\end{figure*}
In Fig.~\ref{fig:cumulative} we present the cumulative number of muon and antimuon neutrino events expected from each source with $N_{\rm fl}$ X-ray flares. Motivated by the previous discussion, we only show results for flares with $\Delta t <30$~d (happening after 54562~MJD). Each curve is normalized to the total expected number of neutrino events. We find that the majority of sources exhibits less than 10 flares contributing to the neutrino signal over the course of the \swift \, coverage. The abrupt increase in the cumulative neutrino number found for a few sources, including Mkn~421 (index 37), occurs at the first flare happening after the starting date of IceCube with the IC40 configuration.
Inspection of the cumulative curves of 1ES~1959+650 (index 6) and Mkn~501 (index 38) shows that a similar total number of flares ($N_{\rm fl}\sim 90$) contributes to the total expected number of neutrinos for each source. However, the gradient of the two curves is very different, suggesting a different temporal behaviour between these two sources. Indeed, as shown in Fig.~\ref{fig:lc}, 1ES~1959+650 appears to have entered a state characterized by higher average X-ray flux and more variability after $57000$~MJD, while Mkn~501 was more active at earlier times $55000-57000$~MJD. These results highlight the importance of regular X-ray monitoring of blazars over long time intervals in making robust predictions of their multi-messenger emission.

We move then to compute the total number of muon and antimuon neutrinos, $\mathcal{N}^{\rm (tot)}_{{\nu}_\mu+\bar{\nu}_\mu}$, expected from each source by summing up the expectations of individual flares. Our results are summarized in Fig.~\ref{fig:barplot} (bottom panel) in the form of a bar plot. Sources are marked by an index as dictated in Table~\ref{tab:sample}. Different colours indicate results obtained after excluding blocks of certain durations (see inset legend for details). No results are reported for sources with no blocks satisfying our flare condition (i.e. $f_{\rm B, 1~{\rm keV}}>\mu+\sigma$, see also Section~\ref{sec:flares} for details). There are a handful of sources whose neutrino signal originates solely from long-duration blocks (see e.g. single coloured bars), which are a result of long gaps between XRT observations. In this case, the reported neutrino signal is likely an overestimation. For the remaining sources of the sample, the true neutrino expectation is bounded from the blue and tan coloured bars, with the latter providing a rather weak upper limit. Only 2 sources in the sample have a total neutrino number larger than one after exclusion of long-duration blocks, namely Mkn~421 (index 37) and Mkn~501 (index 38).
None of them has ever been associated with a high-energy neutrino track event, while Mkn~421 has been reported as a candidate source of a cascade-like neutrino event \citep{padovani_2014}. We will discuss the implications of our findings in Section~\ref{sec:discussion}.

While the bar plot in the bottom panel of Fig.~\ref{fig:barplot} provides a quick-look view of our results, it should not be used on its own to directly compare sources in terms of their neutrino output. The results presented in Fig.~\ref{fig:barplot} strongly depend on the number of XRT observations that is displayed on the top panel (grey bars). In general, sources with more observations tend to have higher predicted neutrino numbers (see, for instance, Mkn~421 (index 37) and Mkn 501 (index 38)). This is due to the fact that flares from sources with poorer temporal coverage are more likely to be missed. Moreover, the number of observations for a given source is correlated with the number of observations belonging to flaring blocks (compare grey and maroon bars in top panel). 
There are however exceptions to this general rule. For instance, GB6J0521+2113 (index 33) and GB6J1159+2914 (index 34) have a comparable number of XRT measurements, but differ in the predicted number of events by $\sim$ two orders of magnitude. This difference can be attributed to differences in the number of flaring states (compare maroon bars of objects 33 and 34 in the top figure panel) and IceCube's effective area. 
Hence, neutrino predictions are also affected by the unique temporal behaviour of each source, the physical parameters describing the flaring region, and the source declination as we demonstrate in the following paragraph.

\subsubsection{Effects of model parameters and source declination}\label{sec:param}
So far we have presented results for fixed values of the magnetic field strength ($B'=10$~G) and Doppler factor ($\mathcal{D}=10$) in all sources. Here, we present the effects of both model parameters on the predictions of the total neutrino number from X-ray flares, and discuss the role of the source declination. 

\begin{figure}
\includegraphics[width=0.47\textwidth]{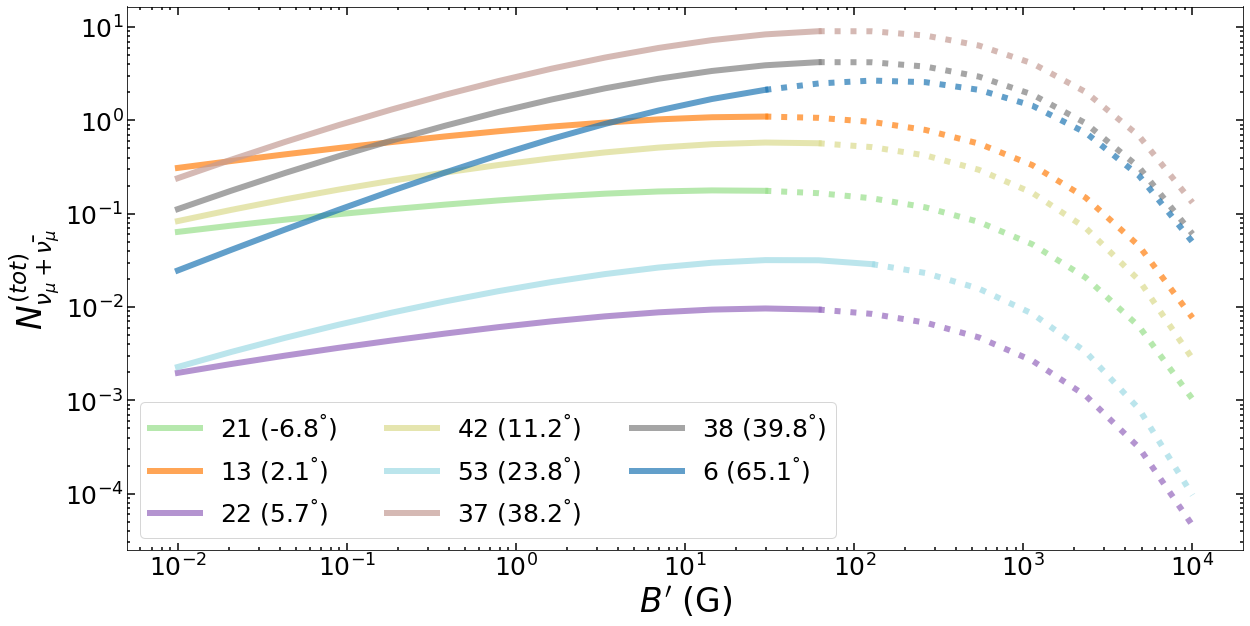}
\includegraphics[width=0.47\textwidth]{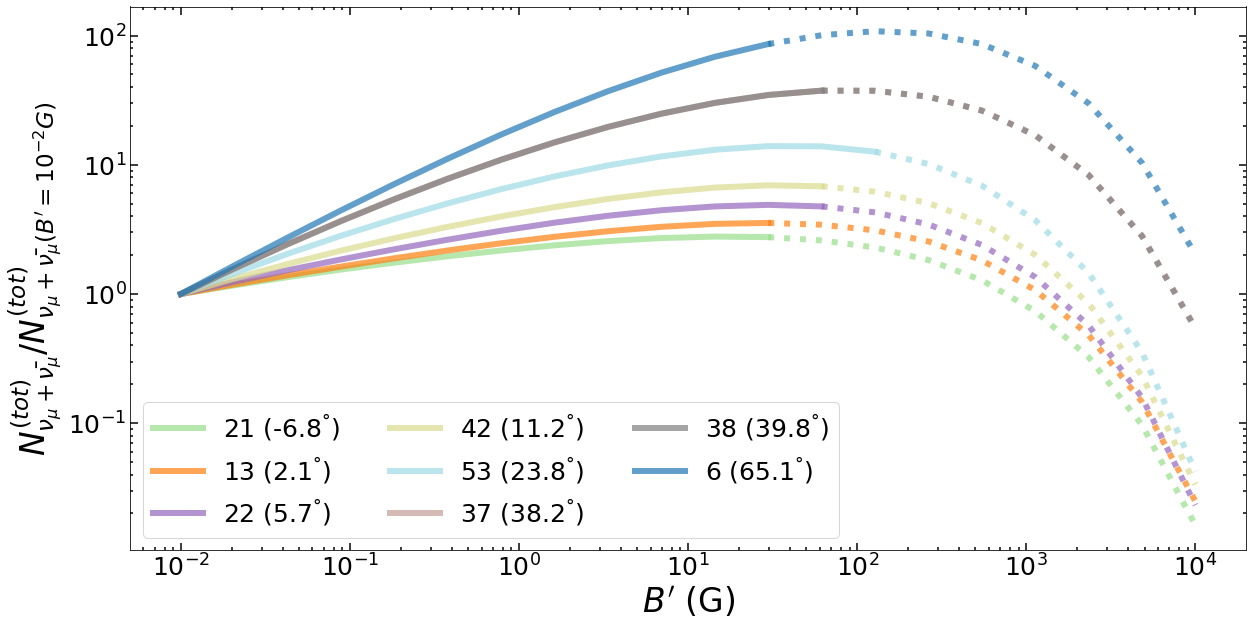}
\caption{\textit{Top panel:} total number of muon and antimuon neutrinos from flares with $\Delta t<30$~d as a function of the magnetic field strength for a few indicative sources (see inset legend; values in the parenthesis indicate declination angles). Other parameters used here are: $\mathcal{D}=10$ and $\varepsilon_{\rm pk}=1$~keV. \textit{Bottom panel:} Same as in the top panel, except that each curve has been normalized to its value at $B'=10^{-2}$~G.}
\label{fig:N_vs_B}
\end{figure}

A higher value of the magnetic field strength $B'$ would lower the proton Lorentz factor $\gamma'_{\rm p}$ needed to produce synchrotron photons of energy $\varepsilon_{\rm pk}$ (see equation~\ref{eq:gp}). For sufficiently strong magnetic fields, it is therefore possible that the proton Lorentz factor drops below the threshold value for pion production on synchrotron photons of the same energy (see equation~\ref{eq:gpth}).  Fig.~\ref{fig:N_vs_B} shows the dependence of the total neutrino number (after excluding long-duration blocks) on $B'$ for some of the sources of our sample whose light curves were presented in Fig.~\ref{fig:lc}. To better illustrate the effects of the magnetic field strength on the total number of events, we adopted a common value for the synchrotron photon energy ($\varepsilon_{\rm pk}=1$~keV) and the Doppler factor ($\mathcal{D}=10$).  Solid lines are used to mark the magnetic field values that satisfy the energy threshold ($\gamma'_{\rm p}> \gamma'_{\rm p,th}$), while dashed lines are used otherwise. 

All curves consist of a power law for sufficiently low magnetic field values, followed by an exponential cutoff for larger values of the magnetic field. The shape of the curves can be understood as follows. Noting that  $\varepsilon_{\nu, \rm c} \propto B^{'-1/2}$ (see equation~\ref{eq:Enu}) and approximating the effective area for neutrino detection with a $\delta$-function centered at the energy of its maximum value, i.e. $A_{\rm eff}=A_0 \delta(\varepsilon_\nu-\varepsilon_{\nu, \rm pk})$, we may write  $\mathcal{N}_{\nu_{\mu}+\bar{\nu}_\mu} \propto A_0 \varepsilon_{\nu, \rm pk}^2 x^{-s+1} e^{-x}$, where $x\equiv \varepsilon_{\nu, \rm pk}/\varepsilon_{\nu, c}$ (see equation~\ref{integral1}). For $x\ll 1$, we recover the power-law dependence on $B'$, i.e. $\mathcal{N}_{\nu_{\mu}+\bar{\nu}_\mu} \propto x^{-s+1}\propto B^{' -(s-1)/2}$, while for $x \gg 1$ we obtain $\mathcal{N}_{\nu_{\mu}+\bar{\nu}_\mu} \propto e^{- a \varepsilon_{\nu, \rm pk} \sqrt{B'}}$ (here $a$ is parameter depending on the Doppler factor and source redshift). Consequently, there is a critical value of the magnetic field, $B'_*$, for each source that maximizes the predicted neutrino number. Under the $\delta$-function approximation for $A_{\rm eff}$, we find 
$B'_* \equiv 10~{\rm G} \, (0.6~{\rm PeV}/\varepsilon_{\nu, \rm pk})^2 (1-s)^2 \mathcal{D}_1 \epsilon_{\rm keV}/(1+z)$. In reality, the dependence of $B'_*$ on  $\varepsilon_{\nu, \rm pk}$ is expected to be weaker, since the IceCube effective area has a broad peak. This critical magnetic field value depends on the source declination through $\varepsilon_{\nu, \rm pk}$, which is a decreasing function of the angle $\delta$. As a result, blazars at lower declination angles obtain their maximum neutrino number for lower values of the magnetic field strength than objects at higher declinations. This effect becomes clearer in the bottom panel of Fig.~\ref{fig:N_vs_B} where each curve is normalized to its value at $B'=10^{-2}$~G. 

\begin{figure}
\includegraphics[width=0.47\textwidth]{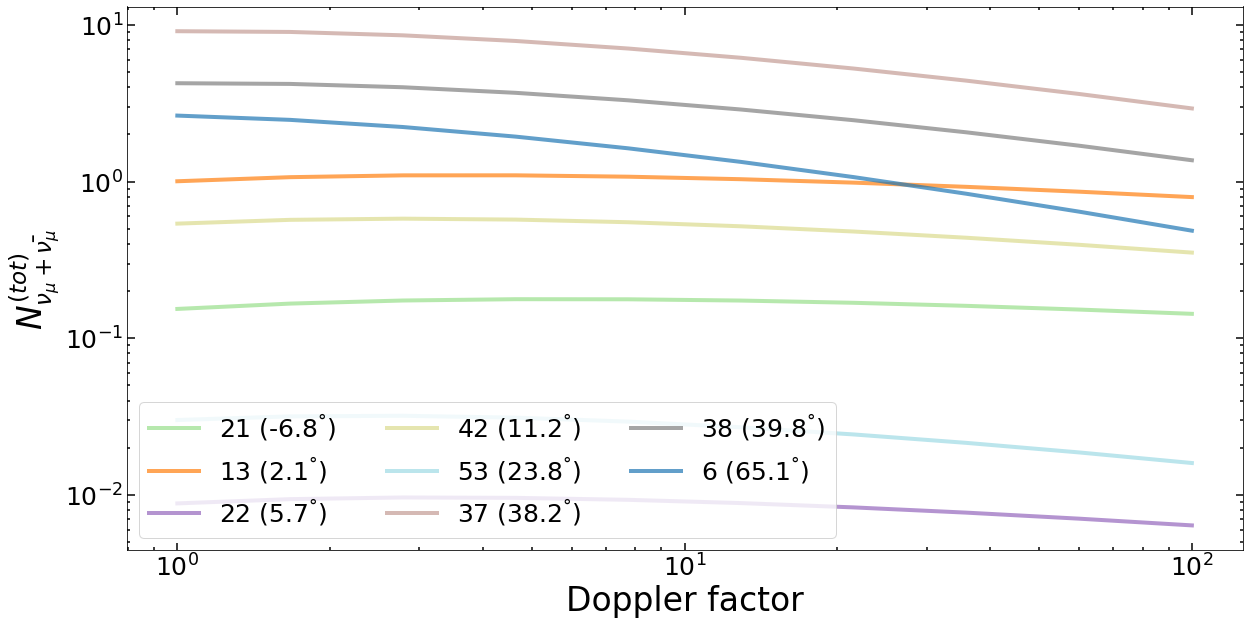}
\caption{Stacked number of muon and antimuon neutrinos from flares with $\Delta t < 30$~d as a function of the Doppler factor for a few indicative sources  (see inset legend; values in the parenthesis indicate declination angles). Other parameters used here are: $B'=10$~G and $\varepsilon_{\rm pk}=1$~keV.}
\label{fig:N_vs_Dop}
\end{figure}
 
Similarly, the predicted total neutrino number depends on the Doppler factor through $\varepsilon_{\nu, \rm c}$. Its effects, however, are less pronounced in the range of values expected for blazar jets, as shown in Fig.~\ref{fig:N_vs_Dop}.

\section{Discussion}\label{sec:discussion}
We have presented predictions for the expected neutrino signal from X-ray blazar flares using a recently proposed theoretical scenario~\citep{2021ApJ...906..131M}. According to it, X-ray flares are powered by synchrotron radiation of intermittently accelerated protons that pion-produce on their own synchrotron photons, thus resulting in a high-energy neutrino flare. Using a sample of 66 blazars that were observed at least 50 times with \swift/XRT, we have computed the number of muon and antimuon neutrinos above 100~TeV expected from X-ray flares over IceCube's livetime. This is the first time (to the best of our knowledge) that \swift/XRT data has been used for this purpose. 
 
The luminosity of the accelerated proton population in the comoving frame, $L'_{\rm p}$, powering an X-ray flare of observed luminosity $L_{\rm X}$ (in the 0.5-10~keV range) is 
\begin{equation} 
L'_{\rm p} \simeq 10^{46} L_{\rm X, 45} R_{16}^{'-1} \mathcal{D}_1^{-7/2} B_1^{'-3/2} \varepsilon_{\rm keV}^{-1/2}(1+z)^{-1/2}~{\rm erg\,s^{-1}}
\label{eq:proton_lum}
\end{equation}
where $R'$ is the radius of the emission region. To derive the equation above, we assumed for simplicity a mono-energetic proton distribution centered at $\gamma'_{\rm p}$ (see equation~\ref{eq:gp}), but these estimates can easily be generalized for a power-law proton distribution. The X-ray luminosity is normalized to $10^{45}$~erg s$^{-1}$, which is close to the median X-ray luminosity of flares (in logarithmic scale) from our sample. The Eddington luminosity of an accreting black hole with mass $M_{\rm BH}$ is  $L_{\rm Edd} = 1.26\times10^{46}~M_{\rm BH}/(10^8 M_{\odot})$~erg s$^{-1}$. The beaming corrected proton luminosity in the observer's frame is $\mathcal{L}_{\rm p} \approx \mathcal{D}^4 L'_{\rm p}/ (2\Gamma^2)$, where a conical jet with half-opening angle of $1/\Gamma$ was assumed. Using equation (\ref{eq:proton_lum}) we find that 
\begin{equation}
\mathcal{L}_{\rm p} \approx 5.7\times 10^{47}~L_{\rm X, 45} R_{16}^{'-1} \mathcal{D}_1^{1/2} \Gamma_1^{-2} B_1^{'-3/2} \varepsilon_{\rm keV}^{-1/2}(1+z)^{-1/2} ~{\rm erg\,s^{-1}}.
\end{equation}
For the default model parameters we find that $\mathcal{L}_{\rm p} \sim 8 \, L_{\rm Edd}$ for $M_{\rm BH}=6.3\times10^8 M_{\odot}$, which assumes the host galaxy to be a typical giant elliptical \citep{Labita_2006}. This black hole mass is close to the mean value  of the black hole masses ($\langle \log( M_{\rm BH}/M_{\odot})=8.6 \rangle$) estimated by \citep{2021ApJS..253...46P} from a large sample consisting of thousands of blazars. Moreover, it is close to the median value of black hole masses recently estimated for a sample of 47 blazars by Padovani et al., 2021, submitted. 
As long as the proton distribution is not a flat power law (i.e. $p>2$) starting from the proton rest mass energy, the energetic requirements of the model are lower than those in other hadronic models for blazar emission~\citep[e.g.][]{Petropoulou_2015, Petropoulou_2016, Liodakis_2020}. The ratio of the (comoving) proton energy density to the magnetic field energy density, $\mathcal{R}=u'_{\rm p}/u'_{\rm B}$, is written as 
\begin{equation}
  \mathcal{R} \simeq 1.5\times10^5 \, L_{\rm X,45} R_{16}^{'-3} B_1^{'-7/2} \mathcal{D}_1^{'-7/2} \varepsilon_{\rm keV}^{-1/2} (1+z)^{-1/2}. 
\end{equation}
The emitting region producing hadronic X-ray flares is therefore far away from energy equipartition between relativistic protons and magnetic fields for default parameter values. However, because of the strong dependence on the magnetic field, equipartition can be reached for $B' > 100$~G (and all other parameters fixed). Thus, strong magnetic fields are more favorable from the energetic point of view, if the size of the flaring region and the magnetic field are not related. 

We adopted a theoretical scenario for transient neutrino production in blazars where the contribution of the hadronic component to the X-ray emission of the source is assumed to be negligible except during flares. To account for the fact that the baseline emission of the source originates from a leptonic component (e.g. electron synchrotron radiation in the case of HBLs), we subtracted from the block flux the mean flux of all X-ray measurements (in the 0.5-10~keV energy range), assuming that the latter is a proxy for the baseline (non-flaring) emission. This ``reduced'' flux was then used for the calculation of the neutrino fluence. In some sources without long-term coverage,  the mean flux may provide an overestimation of the baseline flux. For instance, if 1ES~1959+650 were observed only after $57000$~MJD, its mean flux would be $\sim 1.4$ times higher than the mean flux estimated from  all its measurements (see Fig.~\ref{fig:lc}). Alternatively, one could use the median of all flux measurements as a representative value for the leptonic X-ray flux. We therefore repeated the analysis by subtracting the median of X-ray flux measurements (in the 0.5-10 keV energy range) and found an increase of $\sim 11$ per cent on the median value of $\mathcal{N}_{\nu_{\mu}+\bar{\nu}_{\mu}}$. An increase of $\sim 47$ per cent in the latter quantity was found, when no correction for the baseline emission was made. Hence, a  systematic uncertainty  of $\sim 10-40$ per cent can be assigned to the predicted neutrino numbers to account for the leptonic non-flaring emission.

The X-ray spectrum of FSRQs can be more complicated than in true BL Lac objects due to additional thermal emission components. FSRQs usually exhibit a ``blue bump'' in their low-energy spectra \citep[e.g.][]{1998A&A...340...47P, 2009MNRAS.400.1521J}, which is an indication of emission from the accretion disc. Hence, a fraction of the observed X-ray emission in FSRQs could be related  to thermal radiation from the inner accretion flow~\citep[e.g.][]{2004Sci...306..998G, 2012A&A...541A.160G}. Even though we did not explicitly take into account this component, we assumed that the hadronic population is responsible for emission that exceeds the time-average flux of the source. A more careful analysis of the X-ray flaring spectrum in FSRQs requires detailed modeling of individual sources and lies beyond the scope of this work.

Understanding time variability and flaring states of blazars across the electromagnetic spectrum is a complex subject that is poorly understood. Within the literature even the characterization of flaring or quiescent states is ambiguous \citep[see][and references within]{2009A&A...502..499R}. In X-rays, in particular, the problem is complicated because of the lack of all-sky monitoring surveys that are sensitive enough to provide accurate flux measurements on a daily basis. In the past, instruments like the All-Sky Monitor (ASM) \citep{1996ApJ...469L..33L} on board of the Rossi X-ray Timing Explorer (RXTE) have provided  decade-long data only for a handful of the brightest blazars. However, given the non-imaging nature of the detector, the study of the fainter states remained challenging. For our project we used \swift/XRT data that can provide accurate flux estimates and spectral information for a much larger sample of blazars. Nonetheless, the observations  follow irregular patterns and the observing cadence varies a lot among sources. For example, states with higher fluxes are more likely to be observed multiple consecutive times, whereas low-flux states are more likely to have a few isolated observations as shown in Fig.~\ref{fig:lc}. These issues constitute a ``completeness problem'' that is important to address if one wants to make a meaningful comparison of sources in the sample.  
    
To correct for the incompleteness of light curves in our sample, one could use Mkn 421 or Mkn 501 that have the most well-sampled X-ray light curves.  However, this correction might still introduce errors in the estimation of neutrino counts from different sources, because the properties of X-ray variability are unique among blazars (compare e.g. Mkn 421 and 1ES 1959+650 in Fig~\ref{fig:lc}). For this purpose, we estimate the duty cycle of X-ray flares for each blazar, i.e. the percentage of its life spent on a flaring or a low state. Using RXTE/ASM data, \citet{2009A&A...502..499R} calculated the duty cycle of blazars based on the absolute time the source spends in each flux level. For \swift/XRT data this method cannot be implemented because of the irregular sampling and the large observational gaps. To estimate a duty cycle we therefore need to make some assumptions. First, \swift/XRT sampling is random and the temporal behaviour of a source remains the same when no monitoring data are available. Moreover, we ignore intraday variability, so that each XRT flux measurement is representative of the flux state of the source within that day. The duty cycle can be then defined as the number of \swift/XRT pointings that coincide with a flaring state over the total number of XRT visits, i.e. $d_{\rm fl}\approx N^{(\rm fl)}_{\rm obs}/N_{\rm obs}$.
    
\begin{figure}
    \centering
    \includegraphics[width=\linewidth]{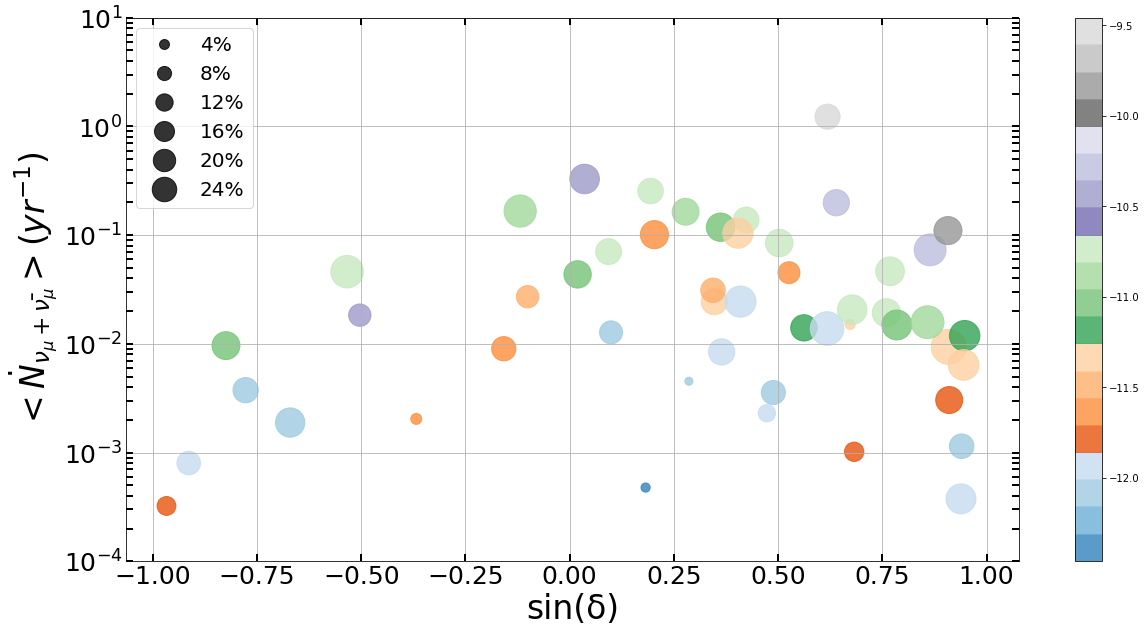}
    \caption{Average yearly muon and antimuon neutrino rate from X-ray flaring blazars as a function of the source declination. The symbol size corresponds to the duty cycle of X-ray flares (see inset legend). Colours indicate the time-average 1~keV flux (in logarithmic scale and in units of erg cm$^{-2}$ s$^{-1}$)}.
    \label{fig:rate}
\end{figure}

For each source in the sample,  we used Bayesian blocks to estimate the duration of the flaring states (i.e. $\Delta t$), we computed the neutrino fluence within these time intervals, and total number of expected neutrinos to be detected from all flares $\mathcal{N}^{(\rm tot)}_{\nu_{\mu} + \bar{\nu}_{\mu}}$. The average neutrino rate of a source due to X-ray flares can be then written as
$\langle \mathcal{\dot{N}}_{\nu_{\mu} + \bar{\nu}_{\mu}} \rangle \approx d_{\rm fl} \mathcal{N}^{(\rm tot)}_{\nu_{\mu} + \bar{\nu}_{\mu}}/\sum_{i}^{N_{\rm fl}}\Delta t_i$. This provides a more representative estimate of the expected neutrino emission than $\mathcal{N}^{(\rm tot)}_{\nu_{\mu} + \bar{\nu}_{\mu}}$, as it roughly accounts for differences in X-ray coverage among sources. The duty cycle and the average yearly neutrino rate of the sources in our sample are shown in Fig.~\ref{fig:rate} (see also Table~\ref{tab:sample}). We find no obvious trend between the average X-ray flux and the duty cycle, while higher yearly rates are expected, in general, for sources with higher average X-ray fluxes. The yearly rate also depends on the source declination through the effective area with a maximum close to 0 deg. Being the brightest X-ray source (on average) in the sample, Mkn~421 has also the highest yearly rate despite its large declination.

So far, only one cascade-like neutrino event (with a median angular error of 16.5 deg) was tentatively associated with Mkn~421~\citep{padovani_2014}. At the time of the neutrino arrival (55685.66~MJD), the source was not flaring in X-rays and the flux was close to its minimum value as shown in Fig.~\ref{fig:lc}. If our model for neutrino production in flaring blazars is true, the lack of high-energy neutrinos from the direction of Mkn~421 \citep{2019EPJC...79..234A} can be explained in two ways: only a fraction of X-ray flares has hadronic origin ($\lesssim 15$~per cent for $B'=10$~G) or $B' \lesssim 0.1$~G in the neutrino emission region (see top panel in Fig.~\ref{fig:N_vs_B}). The source with the second highest total number of events and average yearly rate of events in our sample is 3C~273 (see Table~\ref{tab:sample}), a flat spectrum radio quasar (FSRQ) at redshift $z=0.158$. Excluding flaring blocks with $\Delta t > 30$~d, our model predicts less than 1 muon neutrino event from all X-ray flares of this source. This is consistent with the lack of neutrino excess above the background from the direction of this source \citep[e.g.][]{2019EPJC...79..234A}. 

Only 3 sources from our sample are positionally consistent with astrophysical muon neutrino track events detected by IceCube, namely TXS~0506+056, 1ES~0229+200 and PKS~1502+106. \swift/XRT data are available close to the arrival time of the high-energy neutrino only in the case of TXS~0506+056 and IC~170922A~\citep{IceCube:2018cha}. Our prediction in terms of muon plus antimuon neutrinos after the neutrino detection is 0.0012 events in $\sim 7$~d. The estimated average rate of muon and antimuon neutrinos from the SED modelling of the 2017 flare is $\dot{\mathcal{N}}_{\nu_{\mu}+\bar{\nu}_{\mu}}\sim 0.1~{\rm yr}^{-1}$~\citep{Keivani2018,2020ApJ...891..115P}. Adopting this rate,
we find 0.0019 events for the same time interval. The two predictions are similar even though the underlying models of electromagnetic emission are different, because the maximum neutrino flux in the model of \cite{Keivani2018} is also limited by the X-ray flux. During the period of the so-called neutrino flare of TXS~0506+056 in 2014/15 \citep{IceCube:2018dnn}, only upper limits from \swift/BAT (15-50~keV) \citep{Reimer_2019} and MAXI (4-10~keV) were available \citep{2020ApJ...891..115P}. Hence, our model cannot be applied to that period. While 1ES~0229+200 has a moderate X-ray flare duty cycle ($\sim 13$ per cent), we identify only two flaring states with our method. Accounting for both, we predict  $\mathcal{N}_{\nu_{\mu}+\bar{\nu}_{\mu}}=0.05 \pm 0.01$ in IceCube's livetime. However, the number drops significantly (by a factor of $\sim 10$), if we remove the flaring state with $\Delta t = 143$~d as being non-physical. It is likely that several other flares were missed due to the irregular pattern of XRT pointings. In the case of PKS~1502+106, we find only one flaring state with $\Delta t \sim 1.7$~d. Based on the available XRT data, we obtain a very low duty cycle for X-ray flares from this source. Hence, the detection of IC~190730A from the direction of PKS~105+106 \citep{2019ATel12967....1T} would be explained as chance coincidence in this model.

We have shown that hadronic X-ray flares can be factories of high-energy neutrinos. The ideal targets for X-ray monitoring in terms of their baseline flux are HSP blazars. This group of blazars has its peak frequency of the synchrotron component at the X-rays. Assuming that every X-ray flaring episode in HSP blazars is generated by a hadronic population, the hadronic X-ray flare, which will be above the X-ray baseline, will also produce a neutrino flare of equal integrated flux. An ideal object for X-ray monitoring is 1ES~0229+200 (extreme synchrotron source) which is in spatial coincidence with an astrophysical muon neutrino track event detected by IceCube \citep{2020MNRAS.497..865G}. The declination of the source ($\delta\simeq20.3$~deg) makes it also suitable for neutrino detection since there are no constraints in terms of the effective area of IceCube for this declination. Continuing with this idea another interesting HSP blazar for X-ray monitoring is PG~1553+113 with moderate X-ray variability (duty cycle $\sim 13$ per cent) and a total number of  $\sim0.6$ predicted muon and antimuon neutrinos. In addition, is one of the blazars with the highest average rate of muon and antimuon neutrinos from flares in our sample with $\langle \dot{\mathcal{N}}_{\nu_{\mu}+\bar{\nu}_{\mu}} \rangle$ $\sim 0.25~{\rm yr}^{-1}$.

The limited sensitivity of the current all-sky surveys (i.e. \swift/BAT, MAXI) allows monitoring of a handful of the brightest blazars.
Moreover, future X-ray all sky monitoring missions will not push beyond the current sensitivity limits. While mission concepts like \textit{STROBE-X} \citep{2019arXiv190303035R} could provide a helping hand, their status is unclear.
From current observatories only \swift/XRT has the flexibility for frequent observations. Thus, continuation and enhancement of \swift/XRT observing campaigns is the only way to obtain meaningful light curves to study flaring variability and constraining the duty cycle of potential neutrino emitting sources.

The scenario of hadronic X-ray flares can be scrutinized with the advent of next-generation neutrino detectors and regular X-ray monitoring of blazars. The combination of larger detection volumes, as in IceCube-Gen2 \citep{IceCubeGen2}, with the location of KM3Net \citep{KM3Net}, Baikal-GVD \citep{BaikalGVD18} and the P-ONE \citep{P-ONE} in the northern hemisphere, will increase the number of high-energy neutrino detections and provide a more uniform coverage of the neutrino sky in terms of sensitivity. Lack of neutrino detections from sources with frequent X-ray flaring activity and high X-ray flare fluences could constrain the magnetic field strength of the flaring region and the duty cycle of hadronic X-ray flares.

\bsp

\section*{Acknowledgements}
We thank the anonymous referee for his/her useful comments and suggestions. SIS and MP acknowledge support from the MERAC Foundation through the project THRILL. MP has been supported by the Deutsche Forschungsgemeinschaft (DFG, German Research Foundation) through Grant Sonderforschungsbereich (Collaborative Research Center) SFB1258 ``Neutrinos and Dark Matter in Astro- and Particle Physics'' (NDM) as a Mercator Fellow. 

GV acknowledges support by NASA Grants 80NSSC21K0213, 80NSSC20K0803 and 80NSSC20K1107 during the duration of the project.

\section*{Data Availability}

All X-ray data used in this work are publicly available 
through the tools and web pages of the Open Universe platform: the VOU-BLazars application \citep{VOU-Blazars} and the interactive table at  the ASI SSDC the Virtual Observatory\footnote{\url{https://openuniverse.asi.it/blazars/swift/}}.

\bibliographystyle{mnras}
\bibliography{paper}

\appendix 
\onecolumn
\section{Our sample}
\begin{table*}[h]
\centering
\caption{Sample of blazars observed more than 50 times with \swift/XRT and model predictions about total number and average yearly rate of muon and antimuon neutrinos expected to be detected by IceCube.}
\begin{adjustbox}{width=1\textwidth}

\begin{threeparttable}
\begin{tabular}{*{10}{c}}
\hline 
Source index & Source name & Dec (deg) & $z$ & Class & $N_{\rm obs}$ & $d_{\rm fl}$ (\%) & $\mathcal{N}^{(\rm tot)}_{\nu_{\mu}+\bar{\nu}_{\mu}}(\Delta t <30)$~d &  $\langle \dot{\mathcal{N}}_{\nu_{\mu}+\bar{\nu}_{\mu}} \rangle (\times 10^{-4}~\rm yr^{-1})$ & $ \dot{\mathcal{N}}^{\rm (atm)}_{\nu_{\mu}+\bar{\nu}_{\mu}}$ ($\times10^{-4}~\rm yr^{-1}$) \\ 
(1) & (2) & (3) & (4) & (5) & (6) & (7) & (8) & (9) & (10)\\
\hline 
0 & 1ES~0033+595 & 59.83 & 0.0860 &			          HSP &260 & 20.4&  $0.066 \pm  0.006$ & $ 732.3   \pm  65.0    $&$ 5.1$\\                          
1 & 1ES~0229+200 & 20.29 & 0.1390 &			          HSP &125 & 13.6&  $0.004 \pm  0.001 $ & $ 244.3   \pm  77.8    $&$ 11.8$\\                    
2 & 1ES~0414+009 & 1.09 & 0.2870 & 			          HSP &60  & 15.0& $0.015 \pm  0.003$ & $ 435.4   \pm  77.6    $&$ 15.9$\\                      
3 & 1ES~0647+250 & 25.05 & 0.2030 &			          HSP &137 &13.1 &  $0.10  \pm  0.01 $ & $ 1382.4  \pm  155.4   $&$ 11.0$\\                       
4 & 1ES~1011+496 & 49.43 & 0.2120 &			          HSP &107 &15.9 &  $0.0140 \pm  0.0009 $ & $ 193.1   \pm  12.9    $&$ 7.3$\\                      
5 & 1ES~1218+304 & 30.18 & 0.1820 &			          HSP &158 &15.2 & $0.048 \pm  0.006 $ & $ 846.4   \pm  108.8   $&$ 9.7$\\                        
6 & 1ES~1959+650 & 65.15 & 0.0470 &			          HSP &717 &15.8 &$0.81  \pm  0.05 $ & $ 1101.4  \pm  68.4    $&$ 4.0$\\                       
7 & 1ES~2344+514 & 51.70 & 0.0440 &			          HSP &352 &17.9 & $0.0186 \pm  0.005$ & $ 149.2   \pm  36.4    $&$ 6.9$\\                       
8 & 1H~0323+342 & 34.18 & 0.0610 & 			          ISP &380 &13.9 &  $0.0197  \pm  0.006$ & $ 140.3   \pm  40.9    $&$ 8.9$\\                         
9 & 1H~1515+660 & 65.42 & 0.7020 & 			          HSP &130 &24.6 & $0.0015  \pm  0.0003 $ & $ 93.8    \pm  18.7    $&$ 4.0$\\                        
10\tnote{**} & 1RXS~J154439.4-112820 &  -11.47& $-$ & HSP &     &$          -	    $      &  $          -	         $ & $          -	        $&$-$\\                   
11 & 2E~1823.3+5649 & 56.85& 0.6640 & 				  LSP &99  &8.1  &  $          -	         $ & $          -	        $&$ 6.4$\\                  
12 & 3C~120& 5.35& 0.0330 &                           LSP &322 &13.4 &  $0.08 \pm  0.02$ & $ 705.2   \pm  168.9   $&$ 14.8$\\                            
13 & 3C~273& 2.05& 0.1580 &                           LSP &599 &17.5 &  $0.80 \pm  0.07$ & $ 3286.1  \pm  296.4   $&$ 15.9$\\                             
14 & 3C~279& -5.79& 0.5360 &                          LSP &888 &10.0 &  $0.09 \pm  0.02$ & $ 271.4   \pm  56.5    $&$ 9.0$\\                          
15 & 3C~271& 69.82&  0.0460 &                         ISP &133 &18.0 &  $0.000020 \pm  0.00002$ & $ 3.8     \pm  3.8     $&$ 3.2$\\                      
16 & 3C~454.3&  16.15& 0.8590 &                       LSP &414 & 14.5&  $0.22  \pm  0.02$ & $ 1640.3  \pm  132.8   $&$ 12.5$\\                        
17 & 3C~66A&  43.04& 0.3406 &                         ISP &255 &7.5  &   $0.0005  \pm  0.0001$ & $ 10.0    \pm  1.9     $&$ 7.5$\\                        
18 & 3FGL~J0730.5-6606& -66.04&  0.1060 &             HSP &72  &11.1 & $0.00030  \pm  0.0002$ & $ 8.0     \pm  4.1     $&$ 0.4$\\             
19\tnote{**} & 3HSP~J022539.1-190035& -19.01&0.4000 & HSP & $-$& $          -	    $      &  $          -	         $ & $          -	        $&$ -$\\             
20\tnote{*} & 3HSP~J123800+263553 & 26.60&  0.2100 &  HSP  &204& $          -	    $      &  $          -	         $ & $          -	        $&$ 10.4$\\             
21 & 4FGL~J1544.3-0649 & -6.82&0.1710 &               HSP &101 &20.8 &  $0.15 \pm  0.03$ & $ 1665.5  \pm  328.6   $&$ 9.0$\\                  
22 & TXS~0506+056 & 5.69& 0.3365 &                    ISP &170 &10.6 &  $0.013 \pm  0.004$ & $ 127.5   \pm  38.8    $&$ 14.8$\\                    
23\tnote{*}& 5BZB~J0700-6610& -66.18& $-$ &           ISP &149 & $          -	    $      &  $          -	         $ & $          -	        $&$ 0.4$\\             
24\tnote{*}& 5BZQ~J0525-4557& -45.97& 1.4790 &        LSP &143 &$          -	    $      &  $          -	         $ & $          -	        $&$ 0.3$\\             
25\tnote{*}& PKS~1130+009 & 0.68& 1.6330 &            LSP &147 & $          -	    $      &  $          -	         $ & $          -	        $&$ 15.9$\\             
26& B3~1633+382 &  38.13& 1.8140 &                    LSP &301 &22.6 &  $0.0036  \pm  0.0009$ & $ 138.4   \pm  33.6    $&$ 8.4$\\                    
27& BL Lac & 42.28& 0.0690 &                          ISP &839 &2.1  & $0.060  \pm  0.004$ & $ 151.3   \pm  10.2    $&$ 8.1$\\                           
28& CTA~102& 11.73& 1.0370 &                          LSP &367 &15.8 &  $0.096  \pm  0.01$ & $ 1010.4  \pm  151.1   $&$ 13.0$\\                            
29& EXO~0706.1+5913& 59.14& 0.1250 &                  HSP &88  &21.6 &  $0.0025  \pm  0.0003$ & $ 158.2   \pm  18.1    $&$ 5.0$\\                   
30& EXO~1811.7+3143& 31.74&  0.1170 &                 HSP &272 &9.6  & $0.042 \pm  0.003$ & $ 451.4   \pm  32.0    $&$ 9.5$\\                   
31& GB6~J0521+2113& 21.22& 0.1080 &                   HSP &117 &16.2 &  $0.07 \pm  0.02$ & $ 1182.6  \pm  280.4   $&$ 11.5$\\                      
32& GB6~J0830+2410& 24.18& 0.9390 &                   LSP &169 &19.5 &  $0.0034  \pm  0.0009$ & $ 244.8   \pm  65.7    $&$ 11.0$\\                    
33& GB6~J0849+5108& 51.14& 0.5830 &                   LSP &137 &1.5  &  $          -	        $  & $          -	        $&$ 6.9$\\                  
34& GB6~J1159+2914& 29.25& 0.7250 &                   ISP &155 &11.6 &  $0.00016 \pm  0.00008$ & $ 35.7    \pm  16.9    $&$ 9.7$\\                   
35& H~1426+428& 42.67& 0.1290 &                       HSP &303 &17.5 &  $0.037  \pm  0.008$ & $ 206.6   \pm  46.3    $&$  8.1$\\                        
36& IZW~187& 50.22&  0.0550 &                         HSP &2083 &16.6& $0.030  \pm  0.003$ & $ 465.4   \pm  47.6    $&$ 6.9$\\                         
37& Mkn~421& 38.21& 0.0300 &                          HSP &2026 &12.7&  $4.2 \pm  0.1$ & $ 12284.7 \pm  345.5   $&$ 8.4$\\                             
38& Mkn~501& 39.76& 0.0300 &                          HSP &1036 &13.8&  $1.37 \pm  0.05$ & $ 1986.8  \pm  79.3    $&$  8.4$\\                           
39& MS~1207.9+3945& 39.49& 0.6170 &                   HSP &557 &17.8 &  $          -	         $ & $          -	        $&$ 8.4$\\                 
40& OJ~287& 20.11& 0.3060 &                           ISP &898 &12.0 &  $0.13 \pm  0.02$ & $ 311.0   \pm  52.2    $&$ 11.8$\\                            
41& ON~231& 28.23& 0.1020 &                           ISP &233 &6.0  &  $0.0034  \pm  0.0006$ & $ 23.0    \pm  4.3     $&$ 10.4$\\                           
42& PG~1553+113 & 11.19& 0.3600 &                     HSP &496 &13.1 &  $0.57 \pm  0.05$ & $ 2543.8  \pm  204.1   $&$ 13.9$\\                        
43& PKS~0208-512& -51.02& 1.003 &                     LSP &304 &12.8 &  $0.00039 \pm  0.00008$ & $ 37.5    \pm  7.8     $&$ 0.3$\\                    
44& PKS~0235+164& 16.62& 0.9400 &                     LSP &373 &1.3  & $0.020 \pm  0.002$ & $ 45.3    \pm  4.5     $&$ 12.4$\\                     
45\tnote{*}& PKS~0506-61 &  -61.16& 1.0930 &          LSP &81  & $          -	    $      &  $          -	         $ & $          -	        $&$ 0.3$\\              
46& PKS~0528+134& 13.53&2.0700 &                      LSP &276 &11.6 & $          -	         $ & $          -	        $&$ 13.0$\\                    
47& PKS~0548-322& -32.27& 0.0690 &                    HSP &321 &21.2 &  $0.0011 \pm  0.0002$ & $ 461.4   \pm  99.5    $&$  0.5$\\                     
48& PKS~0637-752& -75.27& 0.6530 &                    LSP &112 &7.1  &  $0.0001 \pm  0.0001$ & $ 3.2     \pm  4.5     $&$  0.8$\\                   
49& PKS~0921-213& -21.60& 0.0530 &                    ISP &167 &2.4  &  $0.0023  \pm  0.0006$ & $ 20.4    \pm  5.3     $&$ 1.0$\\                    
50& PKS~1222+216& 21.38& 0.4390 &                     ISP &259 &13.9 &  $0.007 \pm  0.003$ & $ 84.3    \pm  38.2    $&$  11.5$\\                     
51\tnote{*}& PKS~1406-076& -7.87& 1.4940 &            LSP &161 & $          -	    $      &  $          -	          $ & $          -	        $&$ 5.6$\\              
52& PKS~1424-41 & -42.11&  1.5220 &                   LSP &227 &17.2 &  $0.00020 \pm  0.00008$ & $ 18.9    \pm  8.0     $&$ 0.3$\\                  
53& PKS~1424+240 & 23.80& 0.6100&                     ISP &120 &18.3 &  $0.031 \pm  0.003$ & $ 1033.1  \pm  94.6    $&$ 11.0$\\                       
54& PKS~1502+106 & 10.49& 1.8390 &                    LSP &120 &1.7  &  $0.00013 \pm  0.00012$ & $ 4.8     \pm  4.6     $&$ 13.9$\\       
\hline 

\end{tabular}
\textit{Notes on columns}. (2): Common or discovery name. (3): Source declination. (4): Source redshifts adopted from \citet{2015Ap&SS.357...75M,2017ATel10491....1C, 2018ApJ...854L..32P,2016A&A...589A..92R, 2018MNRAS.474.3162T, 2019A&A...632A..77C}. (5): Spectral class. (6): Number of XRT observations. (7): Flare duty cycle, defined as the ratio of the number of XRT observations in flaring state and $N_{\rm obs}$. (8): Total number of muon and antimuon neutrinos from flares with $\Delta t <30$~d. (9) Average rate of muon and antimuon neutrinos from flares with $\Delta t <30$~d (defined as $\langle \mathcal{\dot{N}}_{\nu_{\mu} + \bar{\nu}_{\mu}} \rangle \approx d_{\rm fl} \mathcal{N}^{(\rm tot)}_{\nu_{\mu} + \bar{\nu}_{\mu}}/\sum_{i}^{N_{\rm fl}}\Delta t_i $). (10) Yearly rate of atmospheric muon and antimuon neutrinos.
\begin{tablenotes}
\item{*} No flares of Type A or B were identified. \item{**} Objects excluded from the analysis (see Section~\ref{sec:data}).

\end{tablenotes}
\end{threeparttable}
\label{tab:sample}
\end{adjustbox}

\end{table*}

\begin{table*} 
\centering
\contcaption{}
\begin{adjustbox}{width=1\textwidth}

\begin{threeparttable}
\begin{tabular}{cccccccccc}
\hline 
Source index & Source name & Dec (deg) & $z$ & Class & $N_{\rm obs}$ & $d_{\rm fl}$ (\%) & $\mathcal{N}^{(\rm tot)}_{\nu_{\mu}+\bar{\nu}_{\mu}}(\Delta t <30)$~d &  $\langle \dot{\mathcal{N}}_{\nu_{\mu}+\bar{\nu}_{\mu}} \rangle$ (yr$^{-1}$) & $\langle \dot{\mathcal{N}}^{\rm atm}_{\nu_{\mu}+\bar{\nu}_{\mu}} \rangle$  ($\times10^{-4}$yr$^{-1}$) \\
(1) & (2) & (3) & (4) & (5) & (6) & (7) & (8) & (9) & (10) \\
\hline 
 55& PKS~1510-08 & -9.10& 0.3600 &                    ISP &693 &11.8 &  $0.02  \pm 0.01 $ & $ 90.2    \pm  55.8    $&$ 5.6$\\                  
 56& PKS~1622-297 & -29.86& 0.8150 &                  LSP &138 &11.6 & $          -	          $ & $          -	         $&$ 0.6$\\                
 57& PKS~1730-130 & -13.08& 0.9020 &                  LSP &182 &19.8 &  $          -	          $ & $          -	         $&$ 2.8$\\                
 58\tnote{*}& PKS~1830-211 & -21.06& 2.5070 &         LSP &236 & $          -	    $      &  $          -	          $ & $          -	         $&$ 1.2$\\                
 59& PKS~2155-304 &  -30.23& 0.1170 &                 HSP &490 & 10.0&  $0.031   \pm 0.002$ & $ 183.7   \pm  10.0    $&$ 0.6$\\
 60& RXS~J05439-5532 & -55.54& $0.2730$ &                  HSP &90  &15.6 &  $0.008  \pm 0.002$ & $ 93.0    \pm  28.9    $&$ 0.3$\\
 61& S4~0954+658 &  65.57& 0.3670 &                   LSP &199 &14.6 &  $0.0016  \pm 0.0003$ & $ 30.4    \pm  5.3     $&$ 4.0$\\
 62& S4~1749+701 & 70.10& 0.7700 &                    ISP &107 &12.1 &  $0.00017  \pm 0.00006$ & $ 11.4    \pm  4.0     $&$ 3.2$\\
 63& S5~0716+714 & 71.34& 0.3100 &                    ISP &657 &18.6 &  $0.009   \pm 0.001$ & $ 118.8   \pm  15.6    $&$ 3.2$\\
 64& S5~0836+71 & 70.90& 2.2180 &                     LSP &260 & 18.8& $0.009  \pm 0.002$ & $ 64.0    \pm  14.6    $&$ 3.2$\\
 65\tnote{\textdagger} & S5~1803+784 & 78.47 & 0.6800 &                  LSP &154 &8.4  & $<0.000076$ & $<0.10     $&$ 2.7$\\
\hline
\end{tabular}
\textit{Notes on columns}. (2): Common or discovery name. (3): Source declination. (4): Source redshifts adopted from \citet{2015Ap&SS.357...75M,2017ATel10491....1C, 2018ApJ...854L..32P,2016A&A...589A..92R, 2018MNRAS.474.3162T, 2019A&A...632A..77C}. (5): Spectral class. (6): Number of XRT observations. (7): Flare duty cycle, defined as the ratio of the number of XRT observations in flaring state and $N_{\rm obs}$.  (8): Total number of muon and antimuon neutrinos from flares with $\Delta t <30$~d. (9) Average rate of muon and antimuon neutrinos from flares with $\Delta t <30$~d (defined as $\langle \mathcal{\dot{N}}_{\nu_{\mu} + \bar{\nu}_{\mu}} \rangle \approx d_{\rm fl} \mathcal{N}^{(\rm tot)}_{\nu_{\mu} + \bar{\nu}_{\mu}}/\sum_{i}^{N_{\rm fl}}\Delta t_i $). (10) Yearly rate of atmospheric muon and antimuon neutrinos.
\begin{tablenotes}
\item{*} No flares of Type A or B were identified. \item{**} Objects excluded from the analysis (see Section~\ref{sec:data}). \item{\textdagger} Upper limits are quoted whenever the statistical error is larger than the predicted value.
\end{tablenotes}
\end{threeparttable}
\label{tab:sample-cont}
\end{adjustbox}
\end{table*}

\section{1~keV versus 0.5-10~keV light curves}\label{app:lc}
The variability in the X-ray flux is often accompanied by changes in the photon index. The photon index exhibits a complicated behaviour during flaring states. In HSP objects, the photon index usually becomes harder when the source becomes brighter \citep[e.g.][]{2018MNRAS.480.4873A, 2019MNRAS.490.2284M, 2019ApJ...885....8W}. Therefore we expect some differences in the variability properties as derived from the 1~keV and 0.5-10~keV light curves. Moreover, the uncertainty of the flux measurements at 1~keV and 0.5-10~keV, which depends on the data processing, can also differ. In fact, the flux at 1 keV is estimated from the count rate in a narrow energy band (0.3-2 keV) while the 0.5-10 keV flux is calculated using the spectral slope derived from the best-fit spectral model. The latter method introduces larger uncertainties and depends on the fitting of the data at these energies. 

\begin{figure*}
    \centering
    \includegraphics[width=0.9\textwidth]{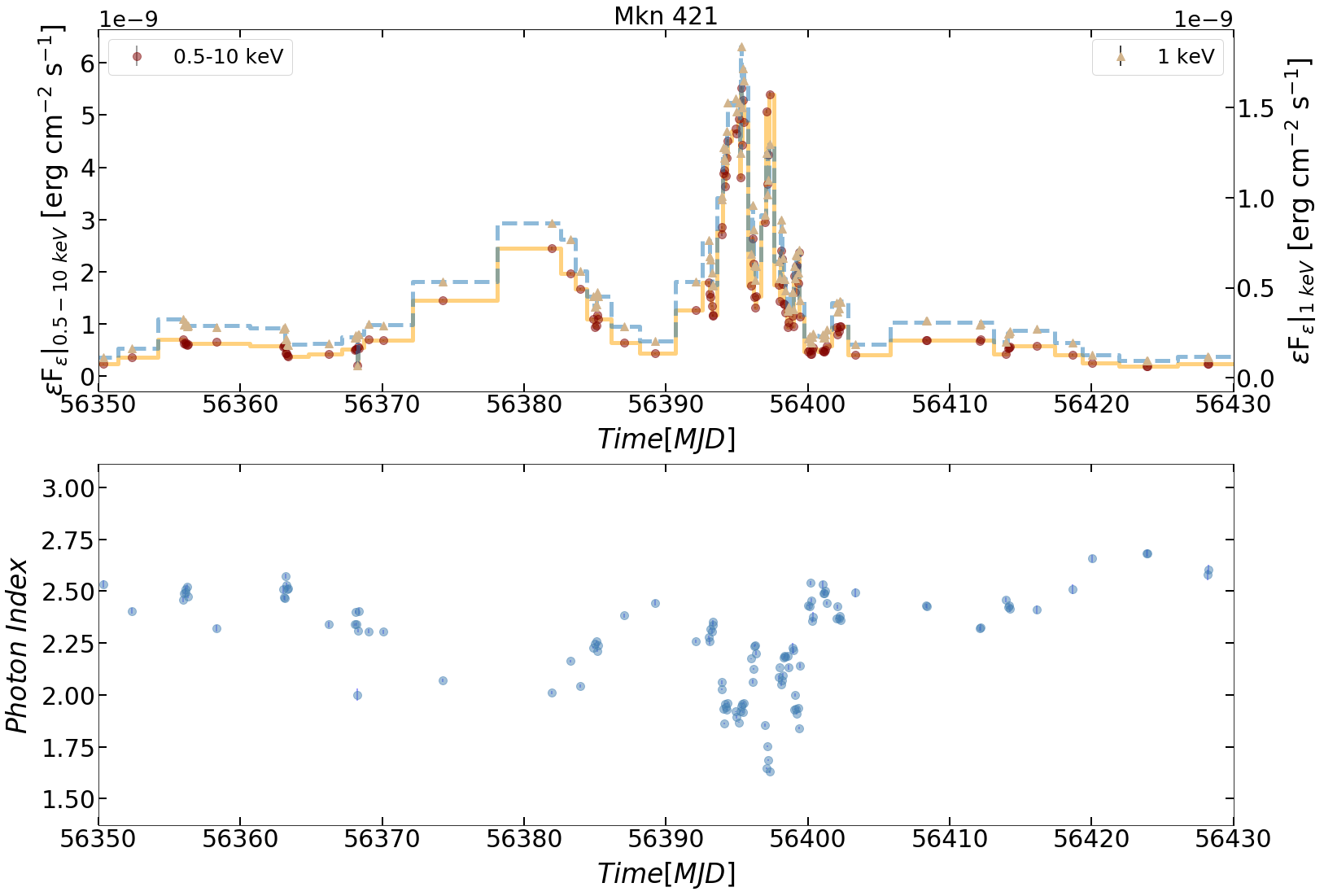}
    \caption{\textit{Top panel:} Segment of the full Mkn~421 light curve at the 1 keV and in the 0.5-10 keV energy range. Solid and dashed lines show respectively the Bayesian block representation of the 0.5-10 keV  and the 1 keV light curves. \textit{Bottom panel:} Photon index of the X-ray spectrum for the same time period. }
    \label{fig:1vs0510keV}
\end{figure*}

Fig.~\ref{fig:1vs0510keV} shows the light curve of Mkn 421 in the 1~keV and 0.5-10~keV energy bands. Overall, we find that fluctuations in flux exhibit the same behaviour. During the time interval of $56390-56400$~MJD,  two flaring events occurred. The spectrum of the first flare centered at 56394~MJD)  has photon index $\Gamma\sim2$. Thus, fluxes in both energy bands would change with time in a similar way. During the following flare, the spectrum was harder with photon index  $\Gamma \sim 1.6$. Hence, the 0.5-10 keV flux differs, and increases by a factor 4.6, from the mean value of all flux measurements (compared to a factor of 3.0 in the case of the 1 keV light curve). In general, for the case of Mkn 421, we found 217 flaring states using the 1 keV light curve from which 185 are Type A and 32 are Type B. Utilizing the 0.5-10~keV light curve we found 178 flaring states out of which 145 are Type A and the rest are Type B using the same criteria for flare classification applied in the 1~keV light curve. Thus, the number of Type B flares remained the same while a smaller number of Type A flares was found in the 0.5-10~keV light curves. Notably, all flares which are reported in the 0.5-10 keV light curve are also identified as flaring states in the 1 keV light curve. 

For completeness, we repeated the flare identification using the 0.5-10~keV light curves of all sources in the sample. In this case, we found a smaller number of flaring states in the sample (723 compared to 967). 
This is a result of larger uncertainties in the 0.5-10~keV flux measurements, which eventually lead to different blocks with different fluxes. Moreover, spectral changes during flares may lead to differences in the flare fluxes at 1~keV and 0.5-10~keV, as illustrated in Fig.~\ref{fig:1vs0510keV}. The percentages of Type A and Type B remain the same in both cases ($84-83\% $ Type A and $16-17\%$ Type B for the 1 keV and 0.5-10 keV light curves, respectively). As a result, the choice of the 0.5-10 keV light curves for the flare identification and classification would reduce the predicted neutrino emission and limit our statistics. 

\section{False-positive rate of Bayesian block algorithm}\label{app:p0}
In this section, we discuss the implications of $p_0$ on the identification of flaring states and neutrino predictions. The choice of $p_0$ in the Bayesian block algorithm is important,  since it is the probability that a change-point reported by the algorithm is truly statistically significant. 

\begin{figure*}
    \centering
    \includegraphics[width=0.85\textwidth]{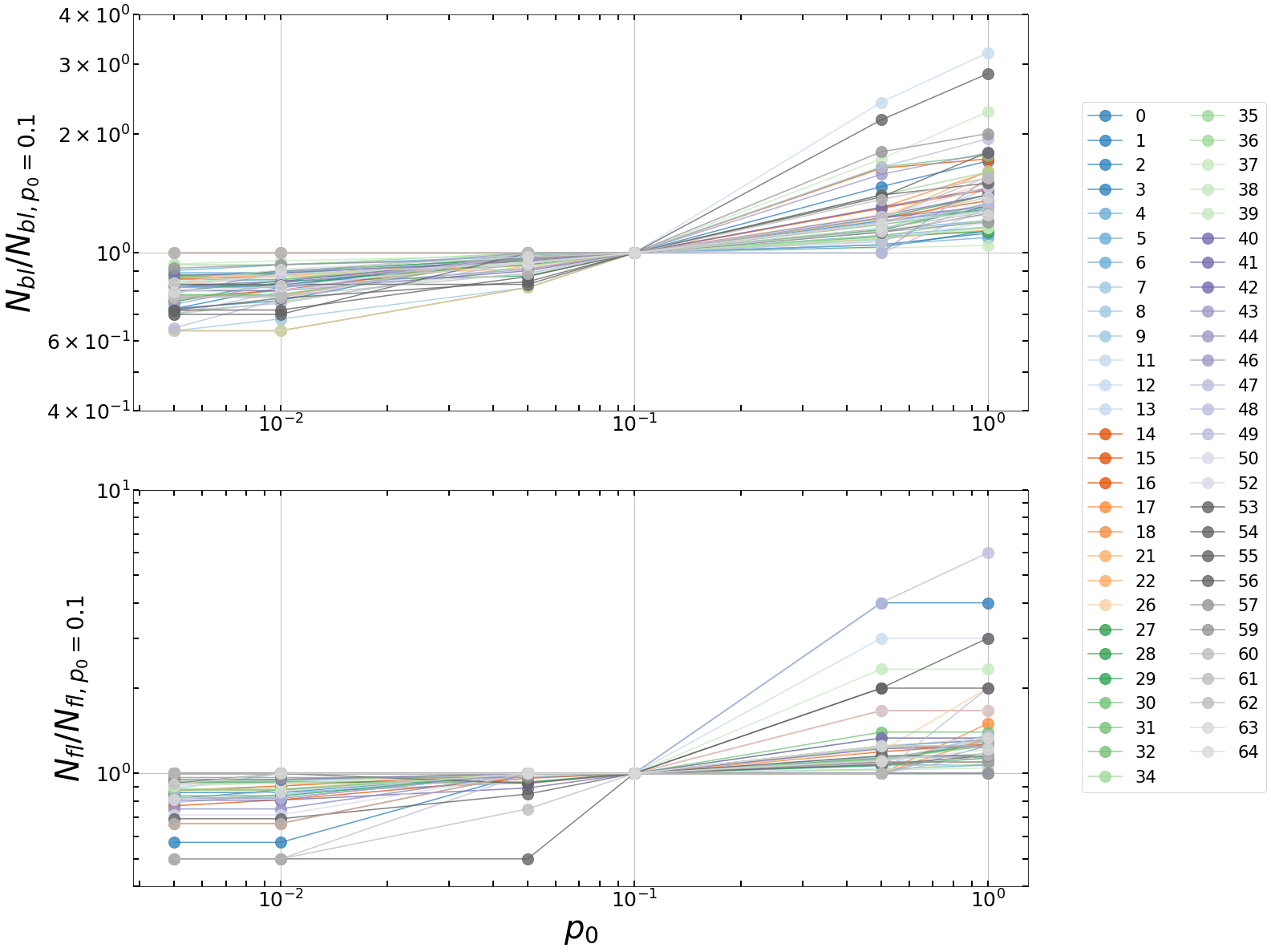}
    \caption{\textit{Bottom panel:} Normalized number of total blocks in all light curves of the sample.   \textit{Top panel:} Normalized number of flaring blocks in all light curves of the sample.}
    \label{fig:NN01}
\end{figure*}

In our analysis we used $p_0=0.1$, while a value of at least $0.05$ is usually adopted in $\gamma$-ray variability studies \citep[e.g.][]{2016A&A...593A..91A, 2019ApJ...877...39M, 2019ApJ...880..103G}. Higher values of probability set a weaker threshold for the identification of statistically significant variations, thus leading to a larger number of blocks detected by the Bayesian block algorithm. This effect is demonstrated in  Fig.~\ref{fig:NN01} where we plot the number of total blocks (top panel) and flaring blocks (bottom panel) as a function of $p_0$ (normalized to their values for $p_0=0.1$) for all sources in the sample. We find that the number of flaring blocks is not very sensitive to the value of $p_0$ for most of the sources in the sample (see clustering of almost horizontal lines around the value of one). Certainly there are a couple of sources where the choice of $p_0$ has a stronger impact on $N_{\rm fl}$, as indicated by  the points with the large scatter. Still, $N_{\rm fl}$ is comparable for $p_0=0.05$ and $p_0=0.1$. More specifically, only 14 sources have a difference in flaring blocks for these specific values of $p_0$, with 12 of them having a difference of only one flaring block. The number of detected blocks, however, depends more strongly on $p_0$ than $N_{\rm fl}$ even for $p_0 \le 0.1$. This result is driven by  blocks with lower fluxes than the adopted threshold for flare definition (i.e. $f_{\rm B} < \mu+\sigma$), which are not used in our analysis. We also note that not every single point of the light curve consists of a block even for $p_0$ values as high as 1. For example, the ratio  $N_{\rm bl}/N_{\rm obs}$ ranges between $\sim 2$ and $\sim 60$ per cent for the sources in our sample for $p_0=0.1$.

\begin{figure*}
    \centering
    \includegraphics[width=0.47\textwidth]{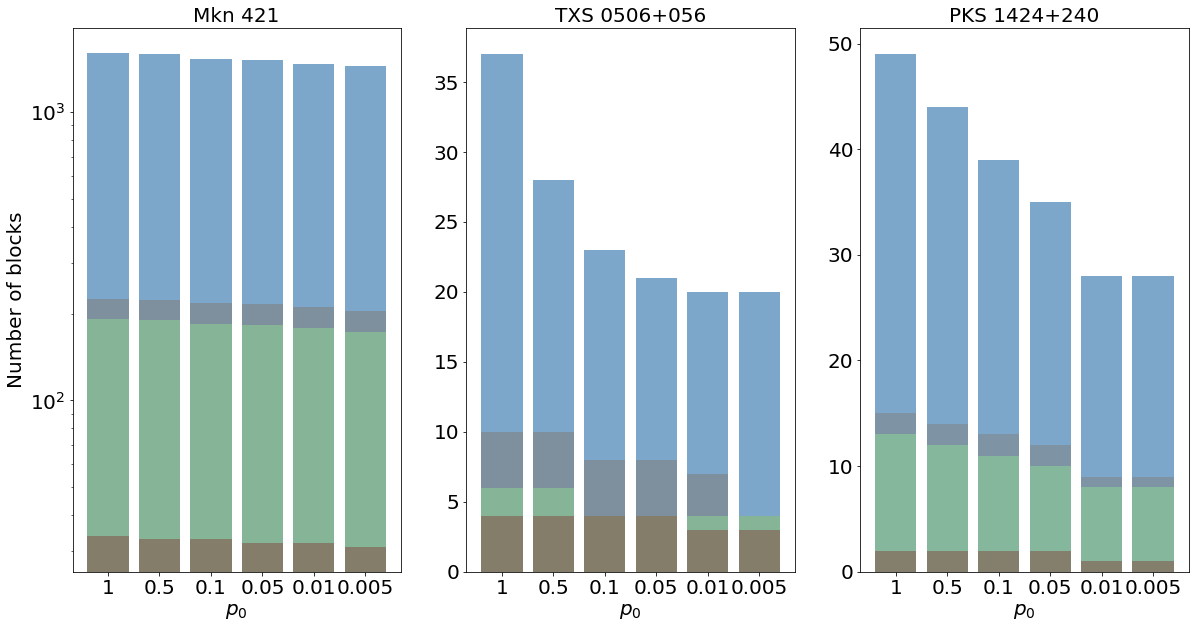}
    \includegraphics[width=0.47\textwidth]{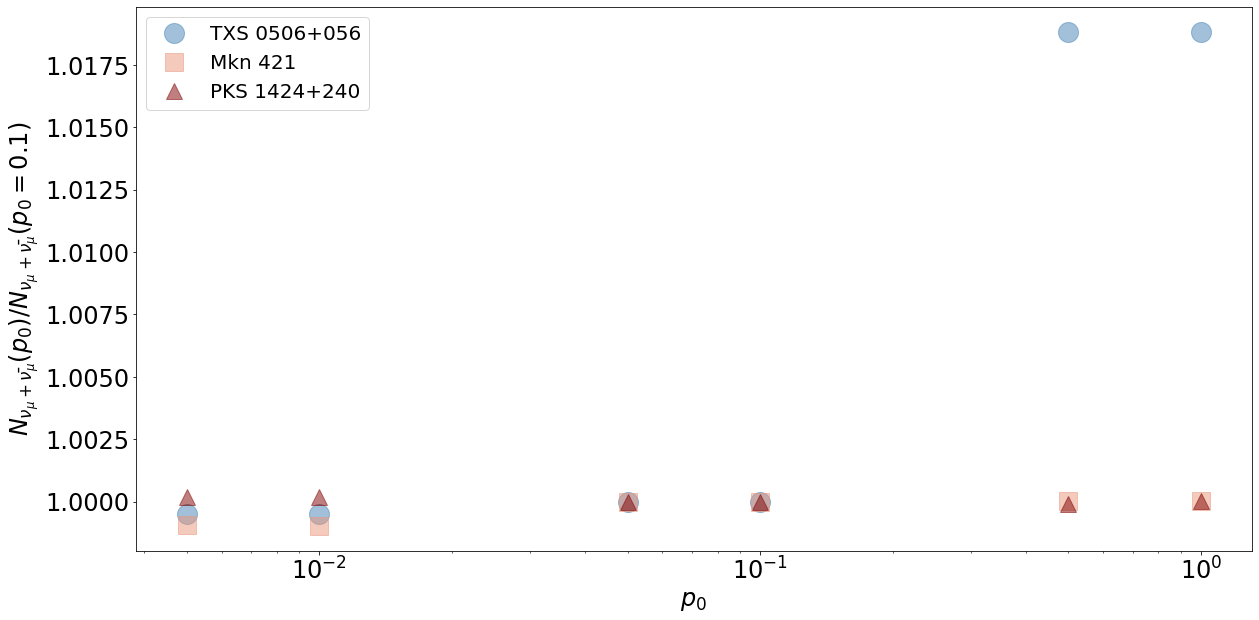}
    \caption{\textit{Left panel:} Number of blocks detected by the Bayesian block algorithm when applied to the 1~keV light curves of three blazars from our sample as a function of the false-positive rate $p_0$. Coloured bars indicate the number of: all blocks (blue), all flares (grey), Type A flares (green) and Type B flares (brown). \textit{Right panel:} Total number of muon and anti-muon neutrino events from X-ray blazar flares as a function of the false-positive rate $p_0$ of the Bayesian block algorithm. The neutrino number is normalized to the value obtained for the nominal value of $p_0=0.1$.}
    \label{fig:Nb_vsp0}
\end{figure*}

We then take a closer look at the impact of $p_0$ on the number of Type A and B  flares as well as on the predicted number of muon neutrinos using three indicative sources from our sample (see Fig.~\ref{fig:Nb_vsp0}). The number of Type A flares, which are characterized by lower fluxes than Type B flares (see definition in Sec.~\ref{sec:flares}), increases for higher values of $p_0$  (see green bars in Fig.~\ref{fig:Nb_vsp0}). This is an expected result since Type A contains more flux measurements inside a flux block compared to Type B flares, which can be interpreted as significant variations by the algorithm for a sufficiently weak limit on $p_0$. A higher value of $p_0$ considers each flux measurement as a unique flaring state inside the light curve and this essentially increases the number of Type A flares. On the contrary, Type B flares many times consist of only one measurement, as it is less likely for higher flux states to last longer. As a result, the increase of $p_0$ does not affect as much these blocks (see brown bars in Fig.~\ref{fig:Nb_vsp0}). A larger value of $p_0$ could be used for objects in the sample that are not ``well-sampled'' or have large uncertainties. In this work, we try to keep our analysis as simple as possible and treat each light curve in the same way. Hence, we select $p_0=0.1$ for all sources in the sample. 
 
Different values of $p_0$ would naturally affect the number of flux blocks and the number of flaring states, but would not have a strong impact on neutrino predictions. Fig.~\ref{fig:Nb_vsp0} (right panel) shows that different values of $p_0$ have almost a zero effect on the predicted  total number of muon and antimuon neutrinos ($\sim1\%$ change in the case of TXS 0506+056). This can be understood as follows. In the left panel of Fig.~\ref{fig:Nb_vsp0} we demonstrated that changes in $p_0$ affect the overall number of blocks, but much less so the number of flaring blocks that contain one flux measurement (Type B flares). In other words, an increase in $p_0$ will divide a block with several flux measurements into blocks with shorter duration containing a smaller number of data points. Thus, the initial information about the fluence of the original flaring block is not lost, but is divided into a larger number of flaring blocks, each having a smaller X-ray fluence. Given that the number of neutrinos from a source depends essentially on the X-ray fluence of the flare, we expect small differences of the total neutrino signal for different values of $p_0$.  

\label{lastpage}
\end{document}